\documentclass[preprint]{iucr}   % 1 column      % DO NOT DELETE THIS LINE

\usepackage{textcomp}
\usepackage[normalem]{ulem}
\usepackage{float}
\usepackage[table]{xcolor}
\usepackage{xcolor}
\usepackage[percent]{overpic}
\usepackage{appendix}

     \journalcode{S}              % Indicate the journal to which submitted
\begin{document}                  % DO NOT DELETE THIS LINE

     %-------------------------------------------------------------------------
     % The introductory (header) part of the paper
     %-------------------------------------------------------------------------

     % The title of the paper. Use \shorttitle to indicate an abbreviated title
     % for use in running heads (you will need to uncomment it).

%\title{\sout{Instrumentation} Development of Multi-Projection X-ray imaging instrumentation for synchrotron and XFEL sources \sout{at EuXFEL}}

\title{Development of crystal optics for Multi-Projection X-ray Imaging for synchrotron and XFEL sources} 

%\shorttitle{Short Title}

     % Authors' names and addresses. Use \cauthor for the main (contact) author.
     % Use \author for all other authors. Use \aff for authors' affiliations.
     % Use lower-case letters in square brackets to link authors to their
     % affiliations; if there is only one affiliation address, remove the [a].

\cauthor[a]{Valerio}{Bellucci}{valerio.bellucci@xfel.eu}

\author[a]{Sarlota}{Birnsteinova} 
\author[a]{Tokushi}{Sato} 
\author[a]{Romain}{Letrun} 
\author[a]{Jayanath C. P.}{Koliyadu} 
\author[a]{Chan}{Kim} 
\author[a]{Gabriele}{Giovanetti} 
\author[a]{Carsten}{Deiter} 
\author[a]{Liubov}{Samoylova} 
\author[a]{Ilia}{Petrov}
\author[a]{Luis}{Lopez Morillo} 
\author[a]{Rita}{Graceffa} 
\author[a]{Luigi}{Adriano} 
\author[b]{Helge}{Huelsen} 
\author[b]{Heiko}{Kollmann}
\author[c]{Thu Nhi}{Tran Calliste} 
\author[d]{Dusan}{Korytar} 
\author[e]{Zdenko}{Zaprazny}
\author[f,g]{Andrea}{Mazzolari} 
\author[f,g]{Marco}{Romagnoni} 
\author[h]{Eleni Myrto}{Asimakopoulou} 
\author[h]{Zisheng}{Yao} 
\author[h]{Yuhe}{Zhang} 
\author[i]{Jozef}{Ulicny} 
\author[j]{Alke}{Meents} 
\author[j,m]{Henry N.}{Chapman} 
\author[a]{Richard}{Bean}
\author[a,k,l]{Adrian}{Mancuso} 
\author[h]{Pablo}{Villanueva-Perez} 
\author[a,j]{Patrik}{Vagovic}

\aff[a]{European XFEL GmbH, Schenefeld, \country{Germany}}
\aff[b]{SmarAct GmbH, Oldenburg, \country{Germany}}
\aff[c]{ESRF - European Synchrotron Radiation Facility, Grenoble, \country{France}}
\aff[d]{Integra TDS Ltd., Piestany, \country{Slovakia}}
\aff[e]{Institute of Electrical Engineering, Bratislava, \country{Slovakia}}
\aff[f,g]{University of Ferrara, Ferrara, \country{Italy}; INFN - Istituto Nazionale di Fisica Nucleare, Ferrara, \country{Italy}}
\aff[h]{Synchrotron Radiation Research and NanoLund, Lund University, \country{Sweden}}
\aff[i]{University of P. J. Safarik, Kosice, \country{Slovakia}}
\aff[j]{Center for Free-Electron Laser Science (CFEL), DESY, Hamburg, \country{Germany}}
\aff[k]{Diamond Light Source Ltd., Harwell Science and Innovation Campus, Didcot, OX11 0DE \country{United Kingdom}}
\aff[l]{Department of Chemistry and Physics, La Trobe Institute for Molecular Science, La Trobe University, Melbourne, Victoria 3086, \country{Australia}}
\aff[m]{University of Hamburg, Hamburg, \country{Germany}}

     % Use \shortauthor to indicate an abbreviated author list for use in
     % running heads (you will need to uncomment it).

%\keyword{keyword}

\maketitle                        % DO NOT DELETE THIS LINE

\begin{synopsis}
%Supply a synopsis of the paper for inclusion in the Table of Contents.
% 1-2 sentences
X-ray Multi-Projection Imaging (XMPI) is an emerging technology that enables the acquisition of millions of 3D images per second, useful for observing rapid, stochastic phenomena previously inaccessible to conventional tomography. This study explores XMPI schemes and optics compatible with synchrotron and XFEL beams, and it experiments with MHz-rate XMPI at the European XFEL.

\end{synopsis}

\begin{abstract}
% 265 words, should be around 250 words for a research paper.
X-ray Multi-Projection Imaging (XMPI) is an emerging technology that allows for the acquisition of millions of 3D images per second in samples opaque to visible light. 
This breakthrough capability enables volumetric observation of fast stochastic phenomena, which were inaccessible due to the lack of a volumetric X-ray imaging probe with kHz to MHz repetition rate.
These include phenomena of industrial and societal relevance such as fractures in solids, propagation of shock waves, laser-based 3D printing, or even fast processes in the biological domain. 
Indeed, the speed of traditional tomography is limited by the shear forces caused by rotation, to a maximum of 1000 Hz in state-of-the-art tomography.
Moreover, the shear forces can disturb the phenomena in observation, in particular with soft samples or sensitive phenomena such as fluid dynamics. 
XMPI is based on splitting an X-ray beam to generate multiple simultaneous views of the sample, therefore eliminating the need for rotation. 
The achievable performances depend on the characteristics of the X-ray source, the detection system, and the X-ray optics used to generate the multiple views.
The increase in power density of the X-ray sources around the world now enables 3D imaging with sampling speeds in the kilohertz range at synchrotrons and megahertz range at X-ray Free-Electron Lasers (XFELs). 
Fast detection systems are already available, and 2D MHz imaging was already demonstrated at synchrotron and XFEL. 
In this work, we explore the properties of X-ray splitter optics and XMPI schemes that are compatible with synchrotron insertion devices and XFEL X-ray beams. 
We describe two possible schemes designed to permit large samples and complex sample environments. 
Then, we present experimental proof of the feasibility of MHz-rate XMPI at the European XFEL.

\end{abstract}

     %-------------------------------------------------------------------------
     % The main body of the paper
     %-------------------------------------------------------------------------
     % Now enter the text of the document in multiple \section's, \subsection's
     % and \subsubsection's as required.

\section{Introduction}
\begin{figure}
\includegraphics[width=1.0\textwidth]{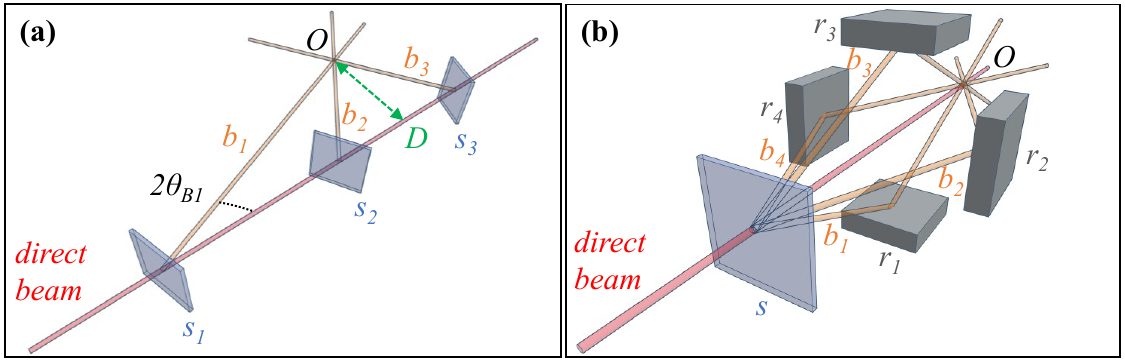}
\caption{Descriptive sketches of the In-Line (a) and In-Parallel (b) multi-projection schemes. (a) Multiple crystal beam splitters ($s_1$, $s_2$, $s_3$) are placed on the direct beam path. Each splitter diffracts a single beamlet ($b_1$, $b_2$, $b_3$) out of the direct beam at an angle equal to twice the Bragg angle ($2 \theta_{B1}$ for the first splitter). The type, position, and orientation of the beam splitters are chosen such that the beamlets converge to a point where the sample object is placed $O$ at a distance $D$ from the direct beam. (b) A single beam splitter $s$ is oriented in the direct beam to excite multiple Bragg diffractions producing several beamlets (4 in the example $b_1$, $b_2$, $b_3$, $b_4$). The beamlets are diffracted by recombiner crystals ($r_1$, $r_2$, $r_3$, $r_4$) toward a common point $O$ on the direct beam path where the sample object is placed.}
\label{fig:schematics}
\end{figure}

Numerous rapid and stochastic phenomena with significant industrial and societal implications take place in materials opaque to visible light. 
These phenomena include the propagation of shock waves \cite{ShockWP:2016, Shockwave:1998}, fractures in stressed solids \cite{Kumar2016StrengthIO, XU2020104165}, laser 3D printing~\cite{synchrotron-laser-3D-printing:2020, diffraction-during-laser-3D-printing:2020}, surface peening~\cite{SOYAMA:2022, Peening-Techniques:2021, Soyama:2023}, and fast biological processes \cite{Hansen2021, Truong2020}. 
Investigating and understanding these complex events is complicated by the absence of a suitable 3D imaging technique with microsecond time resolution. One promising technique for probing such systems is fast 3D X-ray microscopy. 
The current state-of-the-art in fast single projection radiography is primarily limited by the X-ray source's flux and the capabilities of the detector. 
Recent developments have enabled the attainment of Megahertz (MHz) frame rate radiography at synchrotron facilities~\cite{APS:2008, Olbinado:2017} and X-ray Free-Electron Laser (XFEL) sources~\cite{Vagovic:2019}. 
However, when it comes to tomography techniques, the time resolution is primarily constrained by technical considerations such as centrifugal forces, with current rates reaching up to 1 kHz in synchrotron experiments~\cite{Moreno:2021}. 
Centrifugal forces pose a significant technological challenge for the instrumentation and a fundamental challenge for the sample since the shear forces can disrupt the sensitive dynamics under investigation.
Rotation-free kHz and MHz rate 3D X-ray imaging may be attained by X-ray multi-projection imaging (XMPI) schemes. 
These schemes leverage Bragg crystal optics to split the incoming X-ray beam into multiple beamlets, allowing the sample to be examined simultaneously from different angles. 
Subsequently, a 3D representation of the sample is reconstructed using these multiple views, as demonstrated, for instance, by \cite{ONIX:2023}. 
With the centrifugal forces excluded from the system, the maximum acquisition rate would be determined by the luminosity of the setup. 
Therefore, it may be possible to achieve MHz rate 3D X-ray imaging at XFEL sources and kHz rate at synchrotrons.
In this context, the European XFEL is a prime candidate for achieving MHz rate 3D X-ray imaging because of the high flux per pulse and MHz repetition rate of the source.
There have been developments toward 3D kHz imaging at synchrotrons based on XMPI systems \cite{Pablo:2018, Yashiro:2020, DLS:2021-2023}. 
The wavefront of a large white beam can be divided into dozens of small beamlets \cite{Yashiro:2020}. 
This method cannot be used with X-ray beams of small size when imaging a sample of comparable size. 
Therefore, a mm-size XFEL beam would require an amplitude division system to image a mm-size sample. 
The amplitude of a small beam can be divided into multiple, virtually identical monochromatic beamlets by using a single beam-splitter positioned to create multiple beamlets by Bragg diffraction \cite{Pablo:2018}. 
In this case, the coincidence point of the system is in the splitter itself so the sample must be placed as close as possible to the splitter, which limits the size of the sample environment. 
Here we describe two possible schemes \cite{patent:Multi-projection, MHz_XMPI:2023}, designed to overcome those drawbacks and permit larger samples and more complex sample environments, focusing on the crystal optics and related instrumentation. 
These two schemes are referred to here as \textbf{In-Line} (Fig. \ref{fig:schematics}a) and \textbf{In-Parallel} (Fig. \ref{fig:schematics}b) multi-projection geometries. 
Both schemes rely on amplitude splitting, the In-Parallel based on a multi-wave Laue crystal \cite{Pablo:2018} and the In-Line based on a novel in-line configuration of crystal splitters \cite{patent:Multi-projection}.
Both configurations have advantages as the In-Parallel configuration works efficiently with a monochromatic beam (as an XFEL seeded beam with ~1 eV bandwidth) while the In-Line configuration works better with broader band sources ~ 20eV bandwidth and it is tunable in photon energy.
A test experiment of the In-Line scheme at the EuXFEL 1.128 MHz frame-rate and preliminary data for the In-Parallel scheme is reported. 
A further experiment using the In-Line scheme at the EuXFEL has demonstrated two-view imaging of a water droplet collision \cite{MHz_XMPI:2023}. 

This paper includes the simulations that led to the requirements for the optics (Sections \ref{In-Line_Simulations},\ref{In-Parallel_Simulations}), how these optics were developed (Section \ref{Realization-crystals}) and characterized (Appendix \ref{appendix:Topography}). High-precision mechatronics for optics positioning was developed and tested since some of the optics require high accuracy and stability (Appendix \ref{appendix:Mechatronics}). The setups have been experimented on at the Single Particles, Clusters, and Biomolecules and Serial Femtosecond Crystallography (SPB/SFX) instrument of European XFEL (EuXFEL) (Section \ref{In-Line_Demonstration}) recording through two different beamlets simultaneous MHz X-ray radiographs, at the TOMCAT beamline of the Paul Scherrer Institut (PSI) (Section \ref{In-Parallel_Demonstration}) recording intensity curves of the multiple beamlets traversing the sample position, and the ID19 beamline of the European Synchrotron Radiation Facility (ESRF) \cite{ESRF_Myrto:2023}. This work led to the successful demonstration of two-view imaging of two water droplets collision at 1.1 MHz sampling rate at EuXFEL \cite{MHz_XMPI:2023}.

% First image (1)

\section{In-Line Multi-Projection Geometry}
\label{In-Line_Simulations}
The In-Line multi-projection scheme geometry is defined by multiple crystal beam splitters placed sequentially into the path of the incident beam. The parameters, location, and orientation of each crystal splitter are chosen such that a part of the beam is selected and transmitted to a single interaction point where the sample environment is placed. The position $P$ of each splitter is easily calculated 
\begin{equation} 
\label{eq:splitters_positions}
P = D / tan( 2\theta_B ) % I use/reference this equation elsewhere in the text
\end{equation}
where \(D\) is the minimum horizontal distance between the sample and the direct beam, \(\theta_B \) is the Bragg angle of the splitter, and zero is the position closest to the sample.

\subsection{Crystal splitter design simulations } \, 
The scope of a crystal splitter is to divert a large portion of the direct beam into the diffracted branches (beamlets) while absorbing a small fraction of the direct beam so that the beam splitter downstream intercepts an intense beam. The design of a splitter takes into consideration the following parameters: (1) transmission, (2) size of the diffracted beam (field of view), (3) stiffness of the splitter, (4) diffracted intensity, and (5) manufacturing limitations. Here we investigate splitters fabricated in diamond, silicon, and germanium mono-crystals since (a) it is relatively easy to source high-perfection single-crystals of these elements, and (b) these cover a wide range of electron densities, absorption, and diffraction intensity. 

In this study, we take the photon energies of 8 keV, 10 keV, and 15 keV as examples because: 
(i) this range of energy allows studies in mm-size samples with absorption levels from plastic to aluminum; 
(ii) the integrated diffraction efficiency of the splitters is about halved from 8 keV to 15 keV; 
(iii) the angles between the beamlets from the same diffraction planes are also halved from 8 keV to 15 keV, which may decrease the quality of 3D reconstructions \cite{ONIX:2023}.
One might increase the angles between the beamlets (iii) by using diffraction planes of higher order but at the cost of worsening the integrated diffraction efficiency (ii).
Diffraction planes are indicated with the material symbol followed by the Miller index of the plane, e.g. C111 is the diamond diffraction plane (111). 
Here, we design the diamond beam splitter considering the optimization between points (1) to (5) listed above. 

\subsubsection{Transmission:} 
\begin{figure}
\label{fig:transmission}
\includegraphics[width=0.5\textwidth]{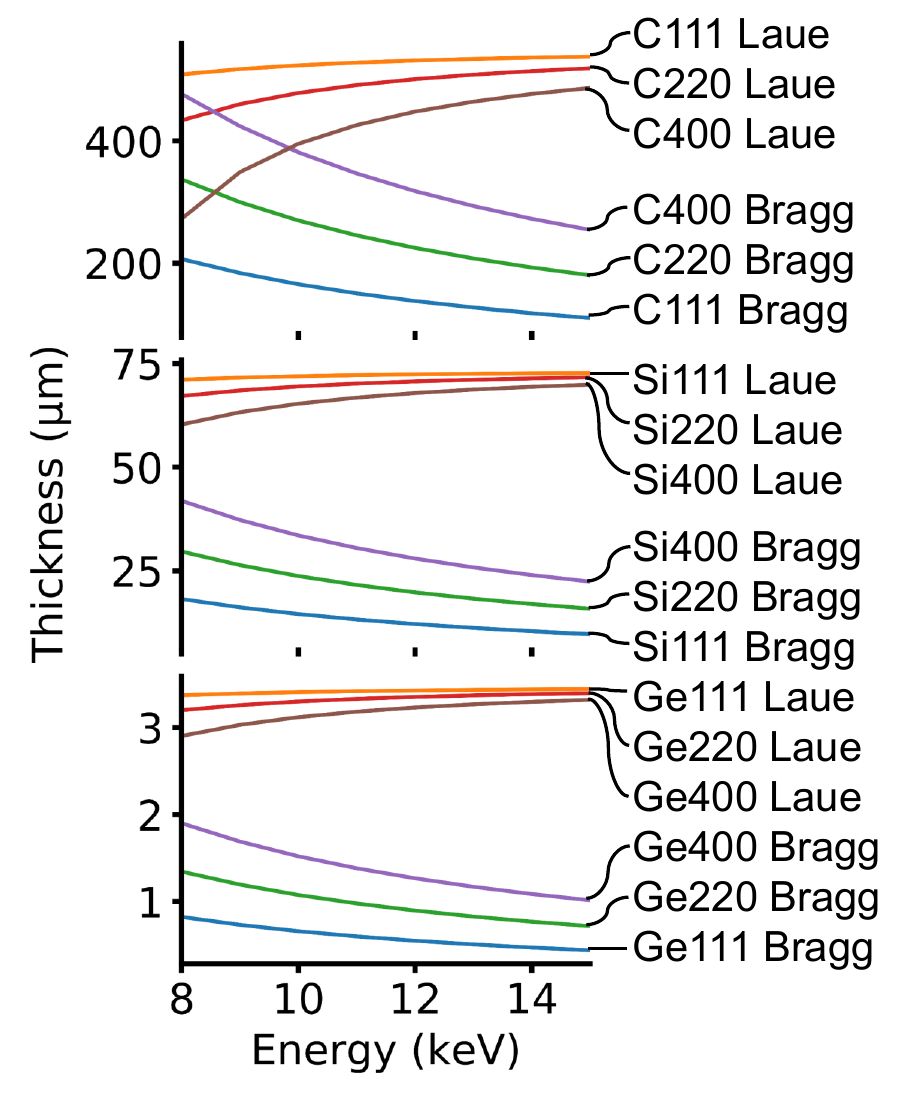}
\caption{Beam splitter thickness vs energy for a resulting 90\% transmitted direct beam, 
when traversing a beam splitter in symmetric Laue or Bragg diffraction geometry, 
for different selected materials (diamond C, silicon Si, germanium Ge) and 
diffraction planes (111), (220), (400) in order of diffraction intensity. 
The selected range of photon energies 8-15 keV is where the In-Line geometry can operate best.  
}
\end{figure}

The transmission of each beam splitter should allow sufficient incident intensity at downstream splitters. 
A threshold of minimum 90\% transmission is chosen here. 
The transmission $I_T$ of the direct beam is calculated as:
The transmission  is calculated as:
\begin{equation}
    \label{eq:Transmission}
    I_T = e^{-\mu L} = e^{-\mu t \ / \ cos(\beta + \theta_B)}
\end{equation}
with $\mu$ the linear absorption coefficient of the material, 
$L$ the length of crystal traversed by the direct beam, 
$t$ the thickness of the splitter, 
$\theta_B$ the Bragg's angle,
$\beta$ the asymmetry angle in Laue geometry, equal to $\beta = 0$ for symmetric Laue geometry and $\beta = \pi / 2 $ for symmetric Bragg geometry. 
In the case of symmetric Laue or Bragg geometry, the traversed length $L$ can be reduced to:
\begin{equation}
\label{eq:thickness-traversed-Bragg} 
{\mathrm{Symmetric \ Bragg \ geometry:}} \ \ L = t \ / \ sin{\theta_B}
\end{equation}
\begin{equation}
\label{eq:thickness-traversed-Laue} 
{\mathrm{Symmetric \ Laue \ geometry:}} \ \ L = t \ / \ cos{\theta_B}
\end{equation}
In the following calculations, we always assume symmetric Bragg and Laue geometry because asymmetric diffraction produces a magnification of the diffracted beam. 
This effect can be used for adjusting the size of the diffracted beam to the field of view of the detector system, as well as for adjusting the passband of the diffraction plane. 
However, this treatment is too specific to the detector system used in the particular setup, therefore it won't be treated here. 
This magnification effect is treated in Appendix \ref{appendix:Magnification} and used in the In-Parallel setup since multiple beam-splitting inherently requires asymmetric diffraction planes. 
A plot of splitter thickness $t$ versus energy at a 90\% transmission condition is represented in Fig. \ref{fig:transmission} for selected materials and diffraction planes.

\subsubsection{Field of view:} 
% description about the optimization, make it for every subsection
The size of the diffracted beams (field of view) should match the maximum sample size that the specific beamline can accept. In this instance, the optimization is carried on for the EuXFEL's SPB/SFX instrument, which has a maximum beam size of 3 mm x 3 mm. In the horizontal scattering geometry, the vertical footprint of the beam on the crystal is equal to the beam height, while the horizontal footprint is a function of the Bragg angle: 
\begin{equation}
\label{eq:footprint-Bragg} 
{\mathrm{Symmetric \ Bragg \ geometry:}} \ \ footprint = beamsize \ / \ sin(\theta_B)
\end{equation}
\begin{equation}
\label{eq:footprint-Laue} 
{\mathrm{Symmetric \ Laue \ geometry:}} \ \ footprint = beamsize \ / \ cos(\theta_B)
\end{equation}
The maximum footprints occur for Bragg (111) diffraction at the highest energy (15~keV). 

\subsubsection{Stiffness:} 
A stiff splitter reduces vibrations that may affect imaging. For a slab of uniform material, the stiffness is proportional to the cube of the thickness while the momenta are proportional to its size \cite{Landau:86}, so the stiffness is maximized by reducing the area while increasing the thickness. 
Therefore, the splitter thickness should be maximized and its area minimized while keeping transmission (1) above 90\%, a large field-of-view (2), a high diffraction efficiency (4). 

\subsubsection{Diffracted Intensity:} 
\begin{figure}
\label{fig:oscillations}
\caption{Integrated diffracted intensity versus beam splitter thickness for a diamond splitter diffracting via its (111) symmetric Laue or Bragg lattice planes, for different selected photon energies 8 keV, 10 keV, and 15 keV. Laue geometry presents symmetry between the diffracted and transmitted beams, which results in oscillations in the diffracted intensity.}
\includegraphics[width=1.0\textwidth]{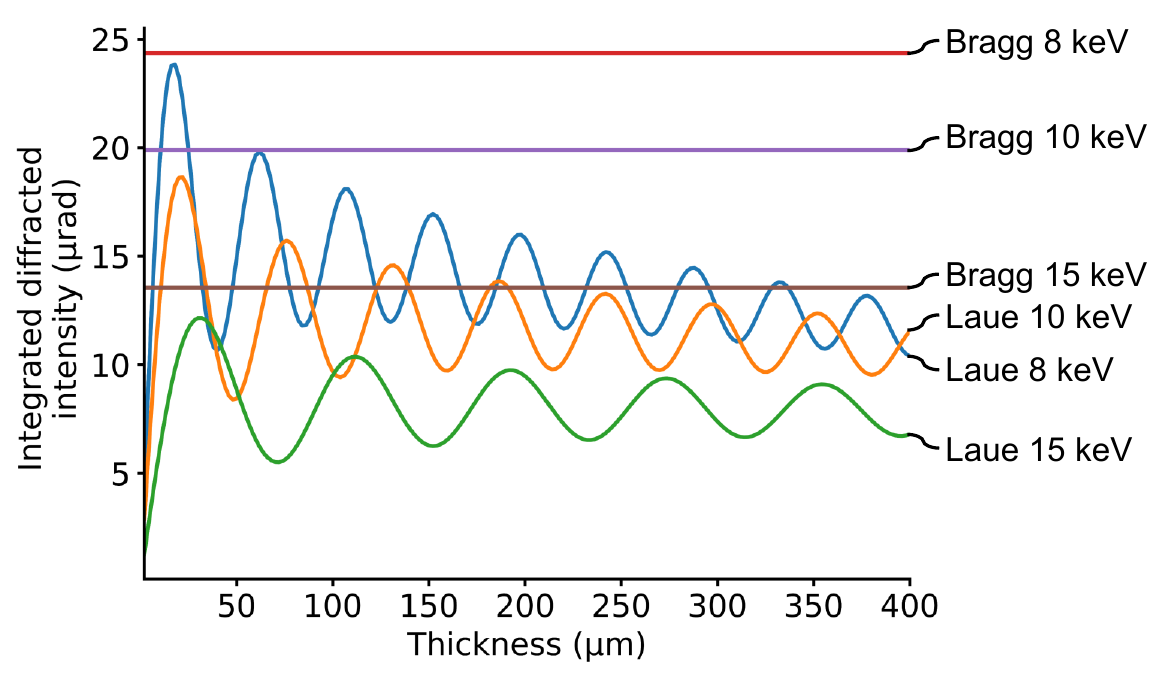}
\end{figure}

A splitter should diffract a large portion of the beam, therefore we optimize the total intensity diffracted by the splitter (integrated diffracted intensity $ I_i^{d} $) versus the thickness of the splitter as per the Dynamical Theory of X-ray Diffraction \cite{Authier:book}. Splitter diffraction in Laue or Bragg geometry follows different functions (Appendix \ref{appendix:Dynamical_Theory_of_X-ray_Diffraction}), so the two cases must be studied separately (Fig. \ref{fig:oscillations}). In both cases, we consider symmetric diffraction geometries. 

The $ I_i^{d} $ function versus the thickness of the Laue splitters follows an oscillatory pattern (Fig. \ref{fig:oscillations}), with the absolute maximum is always reached on the first peak, i.e. the peak with lowest thickness. However, this low thickness may conflict with the technical realization of the splitter (5) and with optimizing its stiffness (3). Moreover, the designed splitter must work for a range of photon energies, but the period of the oscillation changes broadly with the energy, so after the first peak, it is not possible to detect a peak common for the different energies. Therefore, after the first peak, the best option is to wait for the oscillations to stabilize around an average due to the statistical nature of the Pendellösung oscillations.

In the symmetric Bragg case, the $ I_i^{d} $ converges rapidly to an average where oscillations are negligible. On average, the integrated diffracted intensity in Bragg geometry is about 50\% higher than in Laue geometry.

\subsubsection{Manufacturing limitations:} The technical difficulty of realizing crystal splitters increases with thickness $<$ 200 µm.
Silicon splitters of thickness $\sim$~10 µm are commercially available, but such a low thickness allows for warping issues under the heating provoked by an intense X-ray beam \cite{ESRF_Myrto:2023}.
The technology for producing dislocation-free diamonds is currently limited to a 3 mm $\times$ 3 mm clear optic area, i.e. an area free of any dislocation or inclusion \cite{Liubov:2019}. 
Therefore, this is the upper area limit for diamond splitters. 
For silicon and germanium, this technological limit does not exist, so it is possible to accommodate the entire footprint of the beam.

\begin{table}

\begin{tabular}{lccccccc}      % Alignment for each cell: l=left, c=center, r=right
 Diffraction plane           & 111  & 220  & 311  & 400  & 422  & 333  & 440  \\
\hline
 \\
 
  \underline{diamond}  \\
 Laue  \\
 \, thickness (µm)       & 100-120  & 50-100 & 20 or 70 & 15 or 85   & 12 or 100  & 12 or 100  & 30 \\
%  \\
Bragg  \\
 \, thickness (µm)       & 25       & 40     & 50       & 60         & 70         & 70         & 80   \\
 \\
 
\underline{silicon}  \\
 Laue  \\
 \, width (mm)           & 3.1  & 3.3  & 3.4   & 3.7  & 4.2  & 4.5  & 5.1  \\
 \, thickness (µm)       & 10   & 8    & 10-12 & 10   & 10   & 10   & 10   \\
%  \\
Bragg  \\
 \, width (mm)           & 23   & 14   & 12   & 10   & 8    & 8    & 7    \\
 \, thickness (µm)       & 9-10 & 3-5  & 3-5  & 4    & 5    & 8    & 8   \\
  \\

\underline{germanium}  \\
 Laue  \\
 \, width (mm)           & 3.2  & 3.4  & 3.6  & 4.1  & 5.3  & 6.1  & 9.3  \\
 \, thickness (µm)       & 4    & 4    & 5    & 4    & 4.5  & 4.5  & 4.5   \\
%  \\
Bragg  \\
 \, width (mm)           & 18.3 & 11.2 & 9.6  & 8.0  & 6.5  & 6.1  & 5.6    \\
 \, thickness (µm)       & 0.6  & 1    & 2    & 2    & 2.5  & 2.5  & 2.5   \\
  \\
\end{tabular}

\caption{Design simulations for the thickness and size of diamond, silicon, and germanium splitters in the In-Line geometry as from the Dynamic Theory of X-ray Diffraction. The size of clear diamond splitters is limited to 3 mm $\times$ 3 mm by technology, while silicon splitters can be larger. The diffraction geometries chosen are Symmetric Bragg or Laue diffraction. The lattice planes with Miller indices from (111) to (440) are selected. The simulated photon energies are between 8-15 keV. 
}
\label{tab:In-Line_splitters}
\end{table}

\subsection{Diamond, silicon, and germanium splitters} \,
Applying the simulations for the different materials, and lattice planes, and balancing the points (1) to (5), we can obtain the splitter dimensions (Table \ref{tab:In-Line_splitters}).

\subsubsection{Diamond splitters:} \,
The best dimensions simulated for Laue diamond splitters are a thickness of around 100 µm (Table \ref{tab:In-Line_splitters}) according to (1), (3), (4), (5) and an optic area of 3 mm $\times$ 3 mm according to (2), (5). For Bragg splitters, the optimal thickness varies more, spanning from 25 µm for C111 to 80 µm for C440, as it increases with the Miller index.   

\subsubsection{Silicon splitters:} \, 
For Laue silicon splitters, the best dimensions simulated are a thickness of around 10 µm and horizontal size of 5 mm, while for Bragg splitters the thickness ranges from 3 µm to 10 µm with changing Miller index and horizontal size up to 23 mm. The vertical dimension stays fixed at 3 mm to maximize the stiffness (3). The low thickness required is a technological challenge and the first tests of a thin silicon beam splitter are shown in the experimental part of the paper.  

\subsubsection{Germanium splitters:} \,
For a Laue germanium splitter, the best-simulated thickness is about 4 µm, while for a Bragg splitter, it is below 1 µm. Together with the brittle nature of germanium, the low thickness makes this splitter technologically not feasible. Therefore, germanium can be used just as a thick crystal positioned last in the In-Line setup, so its high absorption does not affect other splitters. The size of the optic area is not limited by technology, so it can be up to 19 mm horizontally and 3 mm vertically (2).

\section{In-Parallel Multi-Projection Geometry}
\label{In-Parallel_Simulations}
The In-Parallel multi-projection scheme geometry is defined by a single crystal beam splitter placed on the direct beam path and an ensemble of beam recombiners placed in a conic symmetry around the direct beam path. As for the previously described In-Line geometry, the parameters, location, and orientation of each crystal are chosen such that a part of the beam is diffracted and recombined to a single interaction point where the sample environment is placed.

% Select the splitter lattice family
\subsection{Beam splitter simulations}
\label{subsection:Beam splitter simulations}
The scope of the beam splitter in the In-Parallel geometry is to produce diffracted beamlets in a conical geometry \cite{Pablo:2018}. 
At this scope, the beam splitter was selected between families of lattice planes having cylindrical symmetry (Table \ref{tab:In-Parallel_splitter_families}), which can divide an X-ray beam into identical beamlets by multiple Bragg diffraction. 
Selecting one of these families means fixing the diffraction energy since the Bragg angle is the inclination angle of the plane's family. 
For our setup, we selected a splitter with (100) main face and Laue diffraction planes of the {113} family, having a 17.55° asymmetry angle. 
This geometry is valid both for diamond and silicon splitters since these two elements have the same crystal structure (diamond cubic). 
Nevertheless, diamond and silicon have different lattice parameters, which results in different working energy, respectively 12.4 keV for silicon and 19.1 keV for diamond. 
This particular splitter was selected between the combinations available in Table \ref{tab:In-Parallel_splitter_families} because 
1) the photon energy is compatible with the maximum flux of EuXFEL (8-20 keV), 
2) with a 19.1 keV X-ray beam, it is possible to traverse mm-size aluminum samples, where aluminum alloys are important industrial materials for crack propagation studies, 
3) the 35.1° 2-theta diffraction angle is relatively large, allowing for a compact and portable system, 
4) the {113} family allows for splitting into 8 beams, enabling the expansion of the system to up to 8 beamlets, 
5) both diamond and silicon have low X-ray absorption, and
6) it is technologically possible to realize perfect diamond or silicon crystals of at least mm size. 
Between silicon and diamond, the latter was selected as the best candidate for XFELs due to the lower absorption and larger thermal conductivity, which enable it to better withstand the intense XFEL beam. 
Silicon is better suited for synchrotrons since it provides a larger diffracted intensity in an environment where the thermal load is less critical. 
For lower photon energies, a splitter with (110) main face and diffraction planes of the silicon {220} family at 6.5 keV or diamond {220} family at 9.8 keV is preferable because the (220) diffraction has a larger Darwin's width (Eq. \ref{eq:Darwin_width}) than the (113) diffraction, therefore diffracting a higher flux into the beamlets. 
It is important to point out that the choice of the optimal splitter parameters and working energy changes between the In-Line and In-Parallel geometry because of the different requirements of these two geometries.

\begin{table}

\begin{tabular}{lcccc}      % Alignment for each cell: l=left, c=center, r=right
 Diffraction plane  & Multiplicity      & Asymmetry -   & Energy for        & Energy for  \\
 \,                 & (no. of planes)   & Bragg Angle (°) & Si splitter (keV) & C splitter (keV)\\
\hline
 \\
 
  \underline{Main surface (100)}  \\
 \, \{111\}   & 4     & 35.26     & 3.42      & 5.21 \\
 \, \{113\}*   & 8     & 17.55     & 12.56     & 19.10 \\
 \, \{133\}   & 4     & 13.26     & 21.70     & 33.00 \\
 \, \{224\}   & 8     & 24.09     & 16.30     & 24.80 \\
 \, \{244\}   & 4     & 19.47     & 20.55     & 31.26 \\
 \, \{115\}   & 4     & 11.10     & 30.82     & 46.89 \\
 \, \{135\}   & 4     & 9.73      & 39.95     & 60.78 \\
 \, \{155\}   & 4     & 8.05      & 58.21     & 88.56 \\
 \\

   \underline{Main surface (110)}  \\
 \, \{220\}   & 4     & 30.00     & 6.46      & 9.82 \\
 \, \{113\}   & 6     & 25.24     & 8.88      & 13.52 \\
 \\

   \underline{Main surface (111)}  \\
 \, \{113\}   & 3     & 10.02     & 21.75      & 33.08 \\
 \, \{135\}   & 6     & 43.09     & 9.89       & 15.04 \\
 \,           & 6     & 17.02     & 23.07      & 35.09 \\
 \,           & 6     & 5.6       & 69.20      & 105.27 \\
 
  \\
\end{tabular}

\caption{Selection of beam splitters for the In-Parallel geometry. 
The properties of families of diffraction planes with cylindrical symmetry are studied. 
Multiplicity represents the number of lattice planes in that particular family and symmetry conditions, therefore the number of beamlets that a family can originate. 
Some combination of main surface and diffraction plane can appear at multiple asymmetry angles, ex. the combination with main surface (111) and diffraction plane family \{135\} appear at three different asymmetry angles.  

* The \{113\} family of planes was selected for the In-Parallel setup. 
}
\label{tab:In-Parallel_splitter_families}
\end{table} 
\,

\subsection{Recombiner simulations}
\label{subsection:Recombiner simulations}
Selecting the recombiners (Fig. \ref{fig:schematics}b) also involved iterating through materials and diffraction planes, this time focusing on three points: 

\subsubsection{Angle of view between two opposing beamlets:} The angle of view $\theta_V$ between two opposing beamlets should be as close to  90° as possible to ease 3D reconstruction \cite{ONIX:2023}. 
It can be easily calculated by ray tracing from the Bragg angles of the beam splitter $\theta_{Bs}$ and recombiners $\theta_{Br}$ by: 

\begin{equation} 
\label{eq:In-Parallel_angle_of_view}
\theta_V \, =\, 4 (\theta_{Br} - \theta_{Bs}) 
\end{equation}
as shown in Table \ref{tab:In-Parallel_recombiner_families} for different materials and diffraction planes. 

\begin{table}
\begin{tabular}{l|cc|cc}      % Alignment for each cell: l=left, c=center, r=right
 Recombiners        & Si Recombiners & \, & Ge Recombiners & \,   \\ 
\cline{2-5}
%\hline 
 diffraction        & Si {311}   & C {311}   & Si {311}   & C {311}  \\
 plane              & splitter   & splitter  & splitter   & splitter  \\
\hline
 \{400\}            & 15.10°     & -         & 11.52°     & -        \\
 \{800\}            & 116.43°    & 44.04°    & 106.90°    & 39.06°   \\
 \{12 0 0\}         & -          & 113.07°   & -          & 103.80°  \\
\rowcolor{lightgray}
 \{440\}            & 53.59°     & \cellcolor{white} 8.83°     & 48.12°     & \cellcolor{white} 5.54°    \\
\rowcolor{lightgray}
 \{660\}            & 131.17°    & 51.68°    & 120.88°    & 46.30°   \\
 \rowcolor{lightgray}
 \{880\}            &\cellcolor{white} -          & 99.96°    & \cellcolor{white} 253.17°    & 91.66°   \\
 \{10 10 0\}        & -          & 160.54°   & -          & 146.66°  \\
 \{333\}            & 42.56°     & 2.16°     & 37.66°     & -        \\
 \{444\}            & 85.95°     & 27.63°    & 78.59°     & 23.47°   \\
 \{555\}            & 137.53°    & 54.47°    & 126.15°    & 48.95°   \\
 \{777\}            & -          & 115.52°   & -          & 106.06°  \\
\end{tabular}
\\
\caption{Angle of view between two opposite beamlets ($ \theta_V $, in degrees) for the In-Parallel geometry, considering the 311 diamond or silicon 8-beam splitters and silicon or germanium recombiners. 
The recombiner diffraction planes are looped over the higher orders of the {100}, {110} and {111} planes. 
{110} oriented recombiners were selected (highlighted in light gray) because they present large angles of view with changing splitter, with some angles near 90°.
}
\label{tab:In-Parallel_recombiner_families}
\end{table}

\subsubsection{Diffraction efficiency:}\label{subsubsection:DuMond_diagrams}
\begin{figure}
\includegraphics[width=0.5\textwidth]{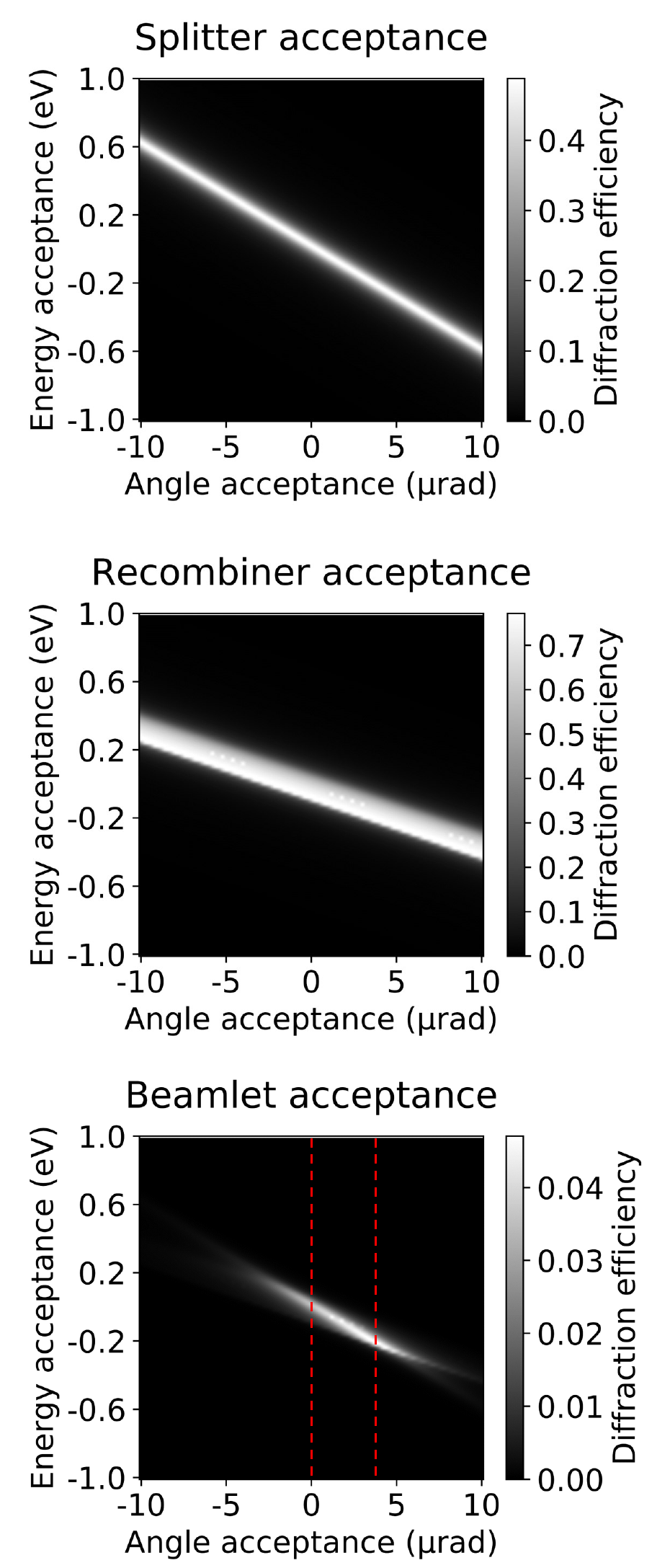}
\label{fig:In-Parallel_DuMond_diagram}
\caption{Simulation of the In-Parallel multi-projection setup acceptance using DuMond diagrams showing diffraction efficiency vs. angle and photon energy. 
The splitter and recombiner acceptances are represented by bands with different widths and inclinations. 
In this example, the splitter is a diamond {113} Laue with asymmetry 17.55°, while the recombiner is a germanium (660) Bragg with asymmetry angle 10° and multiplicity 8. 
The beamlet acceptance is obtained by multiplying the DuMond diagrams of the splitter and recombiner and dividing by the multiplicity of the plane family.
The direct beam is visible in the last graphic as dashed red lines, in this example with photon energy 19.1 keV, chromaticity 20 eV, and angular divergence 4 µrad to simulate the EuXFEL SPB/SFX beam.
This photon energy and the splitter parameters were selected for the reasons listed in the subsections \ref{subsection:Beam splitter simulations} and \ref{subsection:Recombiner simulations}
}  
\end{figure}

Diffraction efficiency is calculated from the dynamical theory of X-ray diffraction (Eq. \ref{eq:IntegratedIntensityBragg}) \cite{Authier:book}. 
The acceptance and diffraction efficiency of a crystal with respect to a range of photon energies and a range of incidence angles can be expressed by a DuMond diagram \cite{DuMond-Davis:1990, Authier:book}. 
Fig.~\ref{fig:In-Parallel_DuMond_diagram} illustrates the DuMond diagrams for the splitter, the recombiner, and the combination of these two elements.  
The integrated diffraction efficiency for each beamlet is obtained by integrating the beamlet acceptance over the chromaticity and divergence of the beam, resulting in \( 0.78 \cdot 10^{-4}\) for the example in Fig. \ref{fig:In-Parallel_DuMond_diagram}. 
The recombiner must be designed in a way that its passband accepts a large fraction of the beam diffracted by the splitter. 
This can be achieved by a wide angular acceptance $\theta_A$: 

\begin{equation} \label{eq:angular_acceptance}
 \theta_A \, = \, 2 \delta_{os}
\end{equation}
where $\delta_{os}$ is the Darwin's width (Eq. \ref{eq:Darwin_width}). 
The acceptance usually increases for heavier materials as it depends on the electron density. 
For the recombiners, transmission is not a design parameter and concerns about thermal load are greatly relaxed since a recombiner intercepts just a beamlet, which contains less than 1\% of the direct beam flux. 
Therefore, we can choose heavier materials, i.e. silicon or germanium versus diamond. 

% Simulations - Recombiners 
Asymmetry can be used for enlarging the acceptance of the recombiners (eq. \ref{eq:Darwin_width}) \cite{Authier:book} while enlarging the physical size of the diffracted beamlet over the diffraction direction by a magnification factor $M$ (Eq. \ref{eq:Magnification}). 
Enlarging the beamlet's physical size can be beneficial since the beamlet was already shrunk due to the asymmetry of the splitter. 
Indeed, the total magnification of the beamlet is attained by multiplying the magnifications produced by the splitter and the recombiner.
Therefore, we can select a recombiner's asymmetry that increases the acceptance while making the shape of the beamlet more symmetric, or similar to the shape of the field of view of the camera. 
For our specific setup, the target camera is the MHz camera Shimadzu HPV-X2. 
Details of the treatment for this case can be found in Appendix \ref{appendix:Magnification}, resulting in a 10° asymmetry angle for the germanium recombiners.

\subsubsection{Ease of alignment and stability of the system after alignment:}\label{subsubsection:Ease of alignment and stability of the system after alignment}
The ease of alignment and the stability of the system following alignment is critical since the beam is diffracted by the splitter and is narrow in chromaticity and divergence, on the order of $10^{-4}$. Therefore, a small misalignment can degrade the diffraction condition. To simplify the alignment, germanium is the most suitable material for the recombiners, having twice the acceptance of silicon and multiple times that of diamond. A grazing asymmetry of 10° further increases the acceptance.

\subsubsection{Selection of the recombiners:} All considerations presented above lead to the selection of germanium recombiners, main face (110) with 10° asymmetry. The germanium {110} family can provide a degree of flexibility at several photon energies (Table~\ref{tab:In-Parallel_recombiner_families}) enabling a range of angles of view including those close to 90°.

\section{Realization of the crystals}
\label{Realization-crystals}
\begin{figure}
\includegraphics[width=1\textwidth]{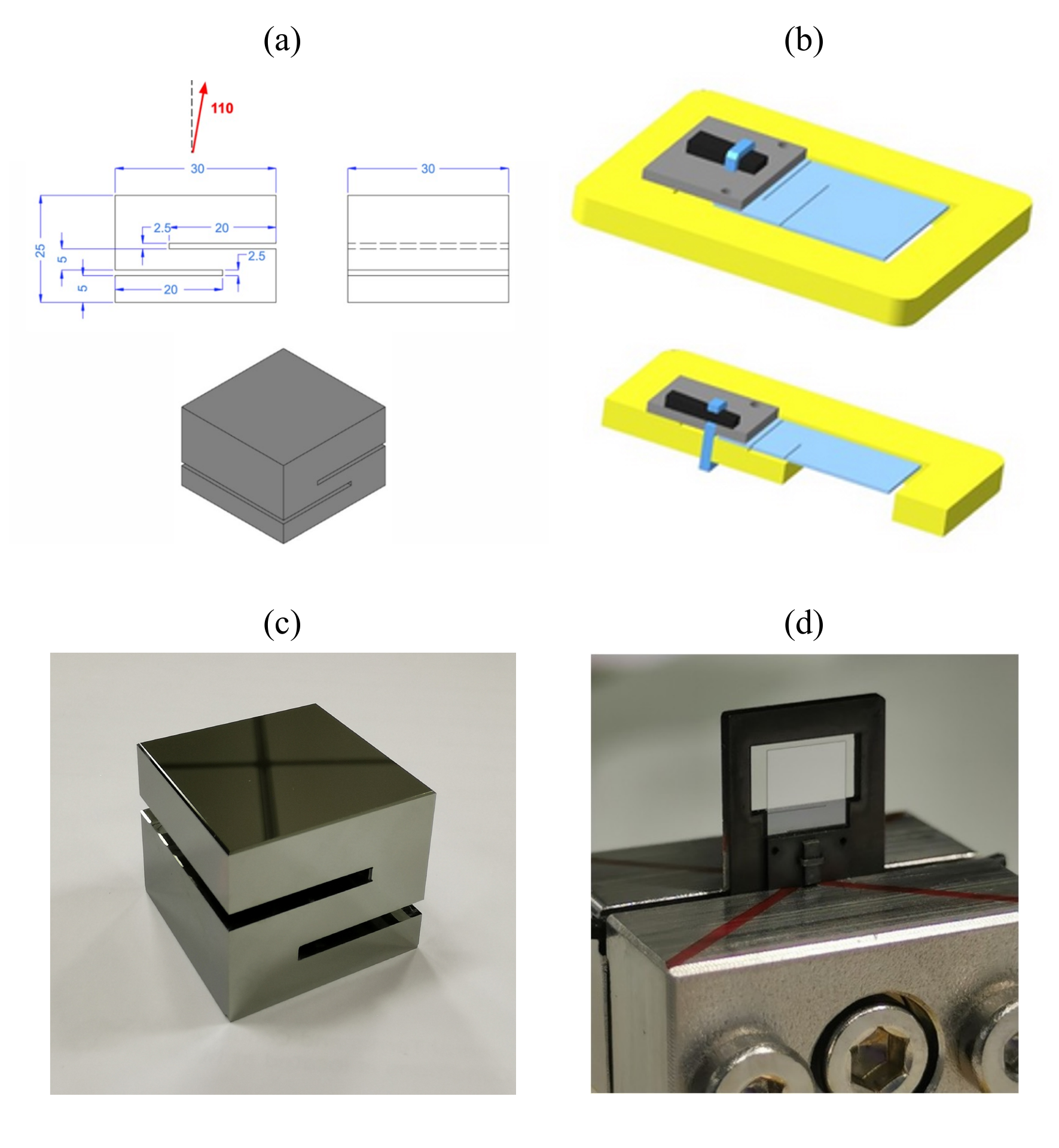}
\label{fig:crystals}
\caption{Crystals used in the multi-projection setup. (a) Drawing and (c) picture of a recombiner. The 2.5 mm large stress-relief cuts are visible, giving an S-shape to the profile of the crystal. (b) Drawing of a diamond beam splitter, light blue being the actual beam splitter, yellow the poly-crystalline frame, and gray graphite used for fixing the two together. Two stress-relief cuts are visible on the base of the beam splitter near the clamping point with the graphite. (d) Picture of a mounted beam-splitter. Both for the recombiners and the splitters, the stress-relief cuts prevent the stress from clamping to propagate to the optic area of the crystal.}  
\end{figure}

The specifications of the crystals were a trade-off between design requirements and technological feasibility. 
The current technology for producing mono-crystalline diamonds (High-Pressure High-Temperature diamonds) allows for reliably producing slabs free of dislocations with an area of 3 mm $\times$ 3 mm or smaller \cite{Ilia:2021}, so this is the maximum size of the optic area. 
The remaining non-perfect part of the slab is used for the strain relief cuts and holding section. 
Diamond crystals are protected by a frame made of poly-crystalline diamond to ease thermal dispersion. 
The splitter is fixed to the frame by the bottom part of the strain relief section to avoid any strain in the optic part (Fig. \ref{fig:crystals}). 

\begin{figure}
\label{fig:311_oscillations}
\includegraphics[width=0.5\textwidth]{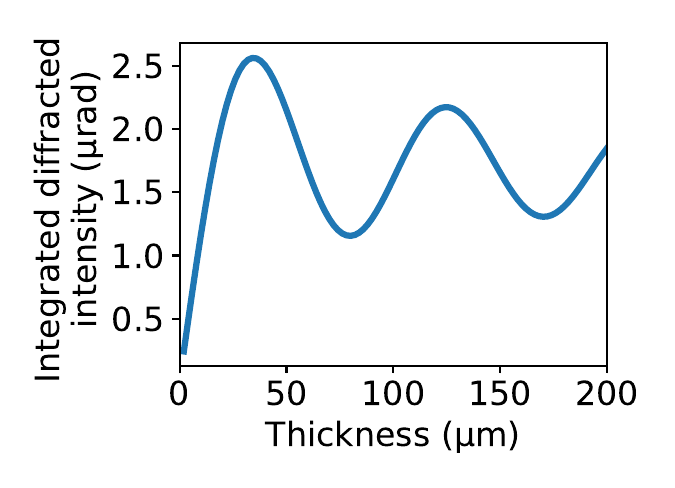}
\caption{Integrated intensity vs thickness for the diamond splitter of the In-Parallel geometry, (113) diamond diffraction plane with 17.5° asymmetry at 19.1 keV.  
}
\end{figure}

The In-Parallel splitter was realized with a 130 µm thickness. 
This value was chosen since it is one of the thicknesses for which the integrated diffracted intensity shows a peak value for the selected {113} diffraction plane family, while the absorption is low, as shown in Fig. \ref{fig:311_oscillations}. 
The thickness at the first intensity peak was not chosen since manufacturing diamond slabs with thicknesses lower than 100 µm presents significant technological challenges. 
The recombiners were made to be as solid and stable as possible while offering a large optic area for diffraction.
Therefore, they were manufactured with an optic area of 30 mm $\times$ 30 mm, a thickness of 25 mm, and with strain-relief cuts 2.5 mm wide, using dislocation-free mono-crystalline germanium. 
All the optic surfaces and their lattice planes are required to be very flat, with residual curvature radius $\geq$ 2.5 km, to accept the low-divergence XFEL beams (i.e. $\geq$ 4 µrad for EuXFEL). 
The roughness and flatness requirements are standard for crystal optics, with roughness (RMS) $\leq$ 1 nm on the scale 10 µm $\times$ 10 µm and the flatness $\leq$ 1 µm over the entire surface.

The quality of the crystals was analyzed by high-resolution monochromatic X-ray diffraction Rocking Curve Imaging technique at the ESRF beamline BM05 (Appendix \ref{appendix:Topography}). 
The diamond splitters performed well during rocking curve imaging, with good crystalline quality through the surface and the bulk. 
Germanium recombiners appear to have a rougher surface, even if the quality is uniform and consistent over the whole sample. 
This rougher surface can be attributed to the brittle structure of germanium and the less-developed finishing  technologies compared to silicon or diamond.

\section{Mechatronics}
Precise 6-axis piezo positioners were developed for the multi-projection systems with SmarAct GmbH 
(Appendix \ref{appendix:Mechatronics:motor_assembly}). 
Indeed, the low acceptance of some of the crystal optics, calls for very precise and stable crystal alignment.
The In-Line geometry has a relatively large tolerance, proportional to the chromaticity of the beam. 
Indeed, if the angle between the direct beam and a splitter changes, the splitter still diffracts X-rays, just with a slightly different energy within the spectrum of the pink beam. 
However, the acceptance of the Bragg angle of the recombiners is particularly small (Section \ref{subsubsection:Ease of alignment and stability of the system after alignment}). 
For this reason, the stability and repeatability of the 6-axis positioners were tested via an interferometric system 
(Appendix \ref{appendix:Mechatronics:Stability}).
The stability measures resulted in an angle drift within 3 µrad over an holding period of 64 hours (Fig. \ref{fig:stability_complete}). 
From the simulations, these conditions are sufficiently stable conditions to align crystalline optics (Section \ref{subsubsection:DuMond_diagrams}). 
The repeatability of the 6-axis positioners was also tested, resulting in a maximum reversal error within 230 nrad, so highly reproducible.

%%%
\section{Experimental demonstration}

\subsection{In-Parallel geometry - Demonstration}
\label{In-Parallel_Demonstration}
\begin{figure}
\label{fig:PSI_setup_1}
\includegraphics[width=0.5\textwidth]{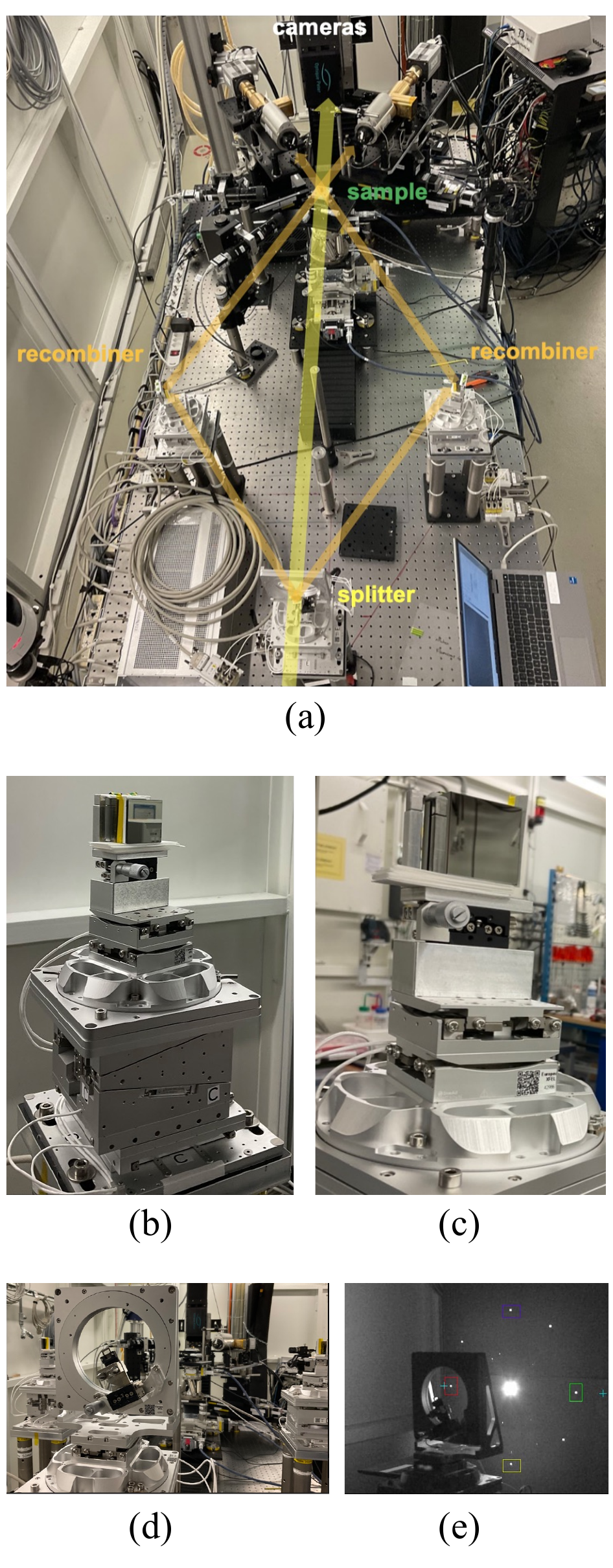}
\\
\caption{Picture of the In-Parallel setup during the experiment at the PSI TOMCAT beamline. 
(a) Overview of the entire setup. (b,c) Horizontal recombiners on their 6-axis piezo positioners. 
(d) Diamond splitter mounted on its positioner. 
(e) Diamond splitter in diffraction position with the X-ray beam shining through. 
The direct beam and the 8 diffracted beamlets from the (113) plane are visible on a scintillator screen placed behind the splitter. 
}
\end{figure}

\begin{figure}
\label{fig:PSI_RCs}
\includegraphics[width=1\textwidth]{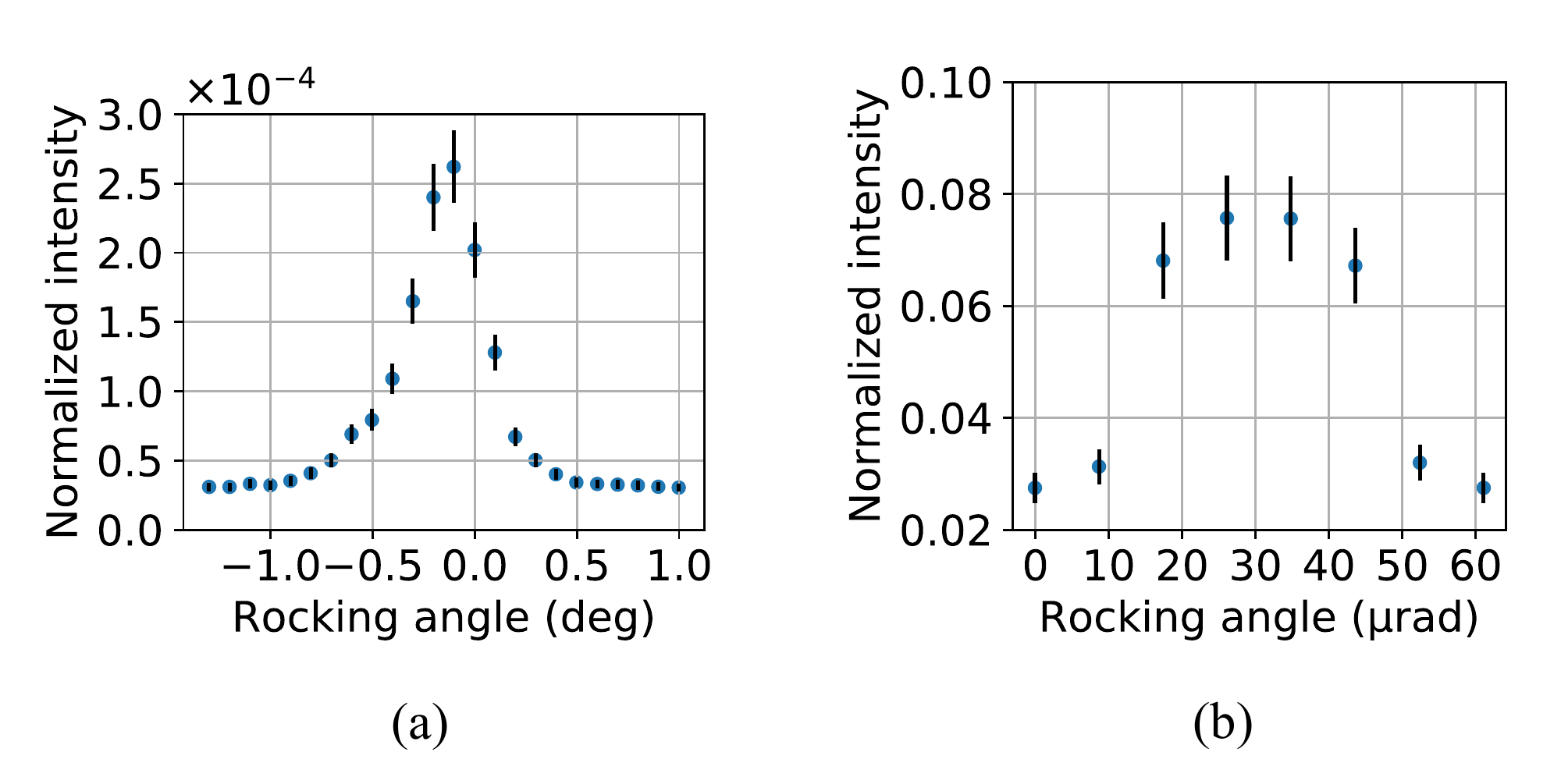}
\\
\caption{Rocking curves of (a) the diamond splitter via one of the beamlets diffracted by the (113) Laue planes with 17.5° asymmetry, (b) the germanium recombiner via diffraction on the (660) Bragg planes with 10° asymmetry. (a) is normalized by the intensity of the direct beam before the splitter, while (b) is normalized by the intensity of the beamlet emerging from the splitter.
}
\end{figure}

% PSI experiment 
The In-Parallel system was tested at the Swiss Light Source synchrotron at the TOMCAT beamline via a pink beam, with a chromaticity of \(10^{-2}\) and an energy of 19.1 keV, to meet the diamond (113) splitter requirements. The splitter was placed to intercept the direct beam and aligned to the position for simultaneous diffraction of 8 beamlets, as shown in Fig. \ref{fig:PSI_setup_1}. The two horizontal positioners were aligned to intercept the beamlets exiting the splitter. By using the (660) diffraction planes of the recombiners, the beamlets were redirected to a common point intercepting the direct beam. In this case, the beam flux provided by the bending magnet beamline was too low to enable the acquisition of good images of a sample. However, we recorded the rocking curves of all the crystals by using a diode. Rocking curves are shown in Fig. \ref{fig:PSI_RCs} for the (660) germanium recombiners and for the (113) diamond splitter. The diffraction efficiency of the splitter is about 70\% of what we expected from the simulations, with \(2.6 \cdot 10^{-4}\) measured versus \( 3.7 \cdot 10^{-4}\) simulated. This discrepancy is probably due to a larger chromaticity and divergence of the direct beam compared with the simulation. For the recombiners, the diffraction efficiency is 0.075 measured versus 0.21 simulated, so about 36\% of the expected value. This larger discrepancy is probably due to the imperfect surface of the recombiners (Fig. \ref{fig:germanium_topography}), which appears rugged when observed at a microscopic level (Appendix \ref{appendix:Topography}). 

\subsection{In-Line geometry - Demonstration}
\label{In-Line_Demonstration}
The In-Line geometry was tested at the SPB/SFX instrument of European XFEL \cite{SPB:2013, Mancuso:2019}. The photon energy is set to 10 keV, with 10 trains per second, each train containing a number of X-ray pulses chosen by the operators between 1 and 300, each pulse delivering on average 3.3 mJ. The spectrum chromaticity is about 20 eV and the divergence is below 4 µrad. The beam size is clipped to 2.4 mm $\times$ 2.4 mm to remove less uniform parts of the beam. The SASE beam instabilities result in a series of artifacts in the images that must be corrected by image processing \cite{Dynamic-flat-field:2015, Birnsteinova:2023}.  

\begin{figure}
\label{fig:EuXFEL_In-Line_picture_1}
\includegraphics[width=1\textwidth]{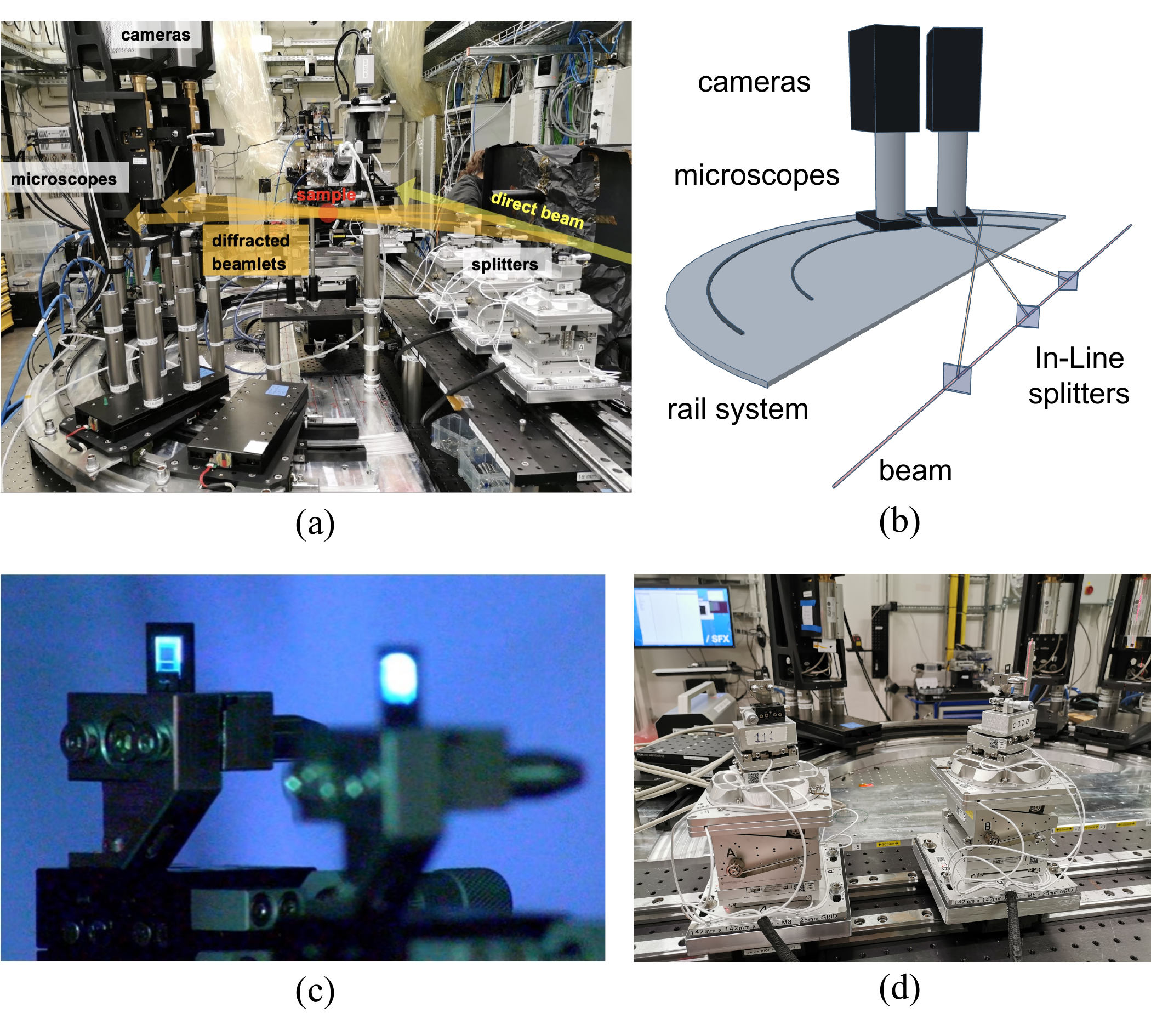}
\caption{(a) Picture of the In-Line setup realized at EuXFEL during the experiment.
(b) Drawing of the mechatronics for the MHz cameras.
(c) Two diamond splitters glowing under the illumination of the EuXFEL beam.
(d) Part of the setup during construction and preliminary testing, with two of the 6-axis crystal positioners in the foreground and the camera positioners in the background. 
}
\end{figure}

The In-Line system is fairly tolerant under a pink beam since slight variations in the crystal orientation would just result in slight variations in the diffracted energy while maintaining the diffraction condition. We first used 110 µm and 130 µm thick diamond splitters via the two most intense Bragg diffraction peaks, (111) and (220), oriented in symmetric Laue geometry to maximize the field of view of the splitters. A Laue symmetric (111) silicon splitter 15 µm thick and 30 mm (H) $\times$ 50 mm (V) in size realized by INFN \cite{Mazzolari_SiMembranes:2014, Germogli_SiMembranes:2014} was also tested to explore the behavior of a silicon beam splitter on the intense beam of EuXFEL. 

A 6-axis Physik Instrumente hexapod was used for positioning a test sample, a metal needle with a thin thread. The sample center was positioned at 300 mm from the direct beam, the minimum distance to avoid collisions between the mechanics and the motors involved. The locations of the splitters are adjusted to diffract the X-ray beamlet to the center of the sample as calculated by Eq. \ref{eq:splitters_positions}, with zero being the position closest to the sample and positive in the direction of the source. Therefore, the splitters were positioned at 181 mm for C220, 428 mm for C111 and 713 mm for Si111, respectively. The direct-beam detector is composed of an Andor Zyla 5.5 sCMOS camera coupled with 5X M Plan Apo infinity corrected Mitutoyo objective looking at a YAG 50 µm thick scintillator via a 45° mirror. 

The splitters are oriented to the Bragg angle and aligned to the maximum in the diffracted intensity via a spectrometer setup \cite{Ilia:2021}. The spectrometer visualizes the energy spectrum of the transmitted beam, showing the spectrum of the direct beam and those parts of this spectrum that were removed by the splitters and transferred to the diffracted beamlets. Looking at these dips in the spectrum, we can align the splitters to diffract the most intense parts of the spectrum, while simultaneously preventing the splitters to superpose, so that each splitter diffracts a different part of the spectrum. In our case, the spectrometer setup is positioned before the direct-beam camera and it is composed of a bent diamond (333) crystal diffracting in Bragg geometry part of the transmitted beam onto an X-ray detector, composed by an Andor Zyla 5.5 sCMOS camera coupled with a 50 µm thick YAG scintillator. The bent crystal offers a different Bragg diffraction angle to every photon energy, so different photon energies are diffracted onto different areas of the camera. Therefore, the image is a direct visualization of the beam spectrum. We used the (333) diffraction because of its narrow energy resolution 0.015 eV when compared to the energy acceptances of the splitters: 0.35 eV for C111 and 0.097 eV for C220.

Each diffracted beamlet passes through the sample and is intercepted by a camera. The mechatronics of the camera imaging and positioning system was developed by SUNA Precision GmbH (Fig. \ref{fig:EuXFEL_In-Line_picture_1}). The main structure consists of a semi-circular rail with the sample position at its center. The cameras move on the rails, so providing a rough alignment between each camera and a beamlet. The fine alignment between each camera and a beamlet is provided by four motors on the camera base. The imaging plane (scintillator position) of each camera is located 500 mm from the sample. A detailed description of the optical system and the hardware integration such as the fast Shimadzu HPV-X2 and Zyla 5.5 cameras are described in \cite{Vagovic:2019}.   
 
The image acquisition by the MHz cameras must be synchronized with the train of X-ray pulses. For this purpose, a MicroTCA (MTCA.4 System, MTCA-6P-PH20x) or a set of Stanford Research DG645 delay generators can be used. The camera frames cannot be perfectly synchronized with the X-ray pulses because the camera's recording speed is specified with a resolution of 10 ns. Our experiment is performed at 1.128 MHz XFEL pulse frequency, so pulses are equally spaced by 886 ns. We, therefore, set the camera speed to 890 ns to approximate the pulse spacing. The mismatch of 4 ns, multiplied by the 128 images in the camera buffer, results in a maximum mismatch of 512 ns inside the train or $\pm$ 256 ns. The YAG:Ce scintillator emission reduces from 100\% to 10\% after about 275 ns following X-ray illumination \cite{Olbinado:2017}. Therefore, we set the camera acquisition window to 590 ns, to prevent capturing two different X-ray pulses in the same camera frame, while keeping the acquisition window as large as possible for capturing a large fraction of each X-ray pulse even at the fringes of the train, when the time mismatch is at its maximum.  

Snapshots from the recorded videos are shown in Fig. \ref{fig:3-images_hair} as stereographic images of the sample, full videos are provided in the supplementary material. The angles between the beamlets are 23.8° between the C220 and C111 beamlets, and 12.2° between the C111 and Si111 beamlets. The C111 beamlet produces images of good quality, reaching contrast-to-noise ratio (CNR) = 14.1 for the detail of the fiber highlighted in Fig. \ref{fig:3-images_hair}a. The C220 beamlet is 4.2 times less intense than the C111 beamlet, so its images have a lower, yet acceptable CNR = 10.1 for the same detail in Fig. \ref{fig:3-images_hair}b. The Si111 beamlet is 3.1 times more intense than C111 resulting in the highest contrast-to-noise ratio, with CNR = 30.9 for the same detail in Fig. \ref{fig:3-images_hair}c. However, the images from the Si111 beamlet present aberrations in the form of duplicated images, i.e. in some of the frames, the object appears to be duplicated. This aberration is caused by the large energy passband of Si111 combined with the SASE spectrum, which is composed of a series of sharp peaks \cite{EuXFEL_spectrum:2020}. When two peaks fall inside the Si111 passband, two beamlets emerge at slightly different angles. As a result, the image appears duplicated. Si111 has the widest passband between the splitters, so it has the highest probability of diffracting two peaks.

\begin{figure}
\label{fig:3-images_hair}
\includegraphics[width=1.0\textwidth]{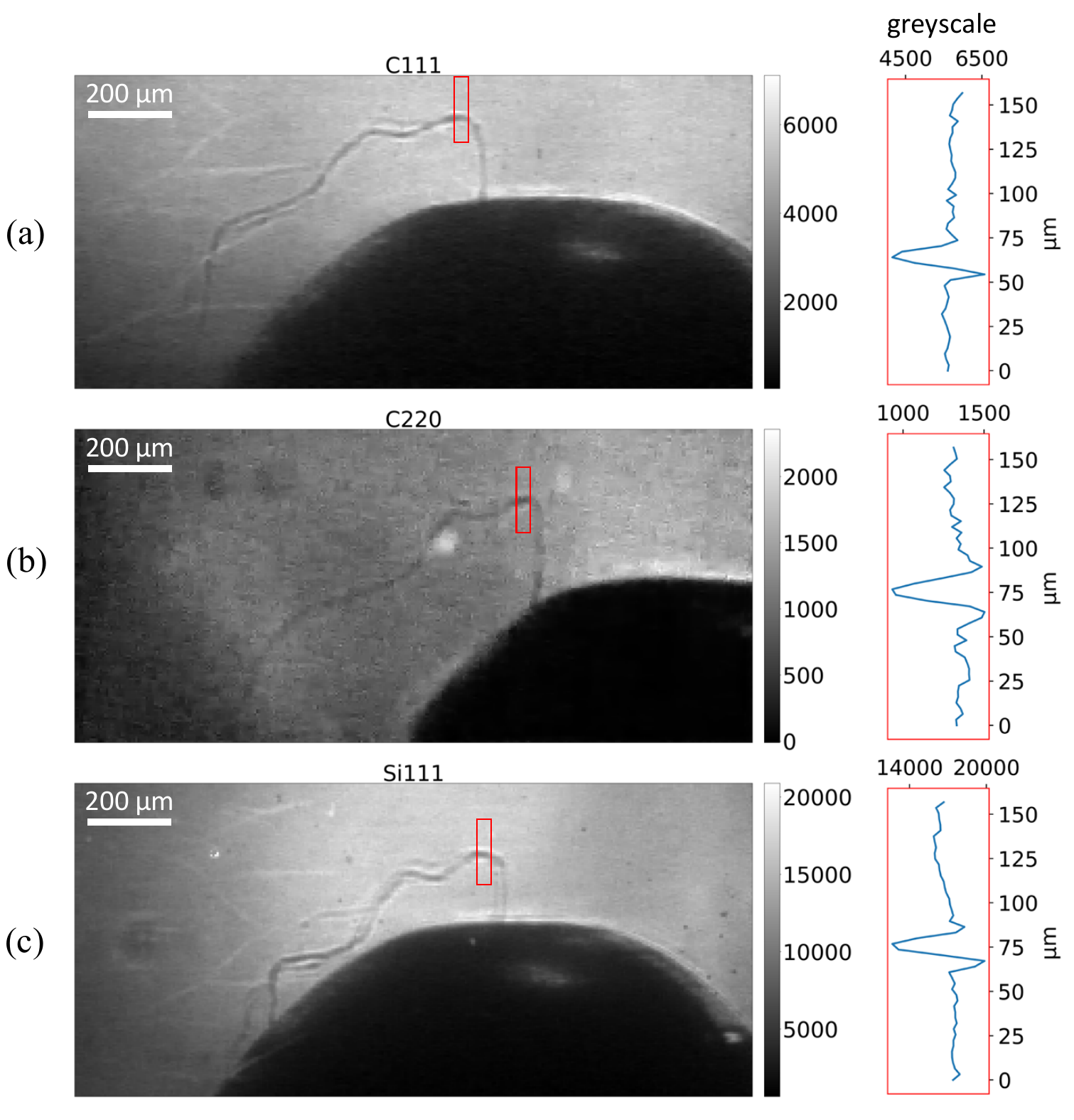}
\caption{Images of the 3 projections captured by the 3 MHz cameras with a single pulse, an acquisition time of 590 ns, a repetition rate of 1.128 MHz, and $10\times$ magnification. The sample is a metal tip with a plastic fiber thread glued on top. In the red box, a section of the image of the fiber is shown with a height of 160 µm, averaged over the width of 32 µm. The projections are from Laue symmetric splitters diffracting via the lattice planes: (a) diamond (111), (b) diamond (220), (c) silicon (111). We  calculate the contrast-to-noise ratio for the detail of the fiber inside the red box, resulting in CNR = 14.1 (a), 10.1 (b), 30.9 (c), respectively.
}
\end{figure}

%%%
\section{Conclusions and Outlook}

In this paper, we developed crystal optics for fast Multi-Projection X-ray microscopy and we demonstrated that, via this instrumentation, it is possible to attain multi-projection X-ray imaging up to a frame rate of 1.128 MHz. 
The presented designs work best at a monochromatic or pink beamline, such as an XFEL beamline with a SASE source. 
This is due to the narrow bandpass of the crystal optics efficiently using a beam with a narrow spectrum. 
We demonstrated the technology enabling multi-projection imaging so that beamlines may offer rotation-free 4D X-ray imaging to their users.  

With this new instrument, beamlines around the world may be able to perform 4D imaging on fast or fragile opaque samples that have never been observed before. 
Future research for developing the multi-projection technology may focus on stable thin membrane-like beam splitters composed of heavier materials to increase the efficiency and luminosity of each projection. 
Further improvement may also come from aligning the diffraction plane of the system in the vertical plane since the horizontal polarization is common in synchrotron or XFEL beams, resulting in small amounts of radiation being diffracted horizontally at Bragg angles near 45°.

     % Appendices appear after the main body of the text. They are prefixed by
     % a single \appendix declaration, and are then structured just like the
     % body text.
\appendix

     %-------------------------------------------------------------------------
     % The back matter of the paper - acknowledgements and references
     %-------------------------------------------------------------------------

     % Acknowledgements come after the appendices
\ack{Acknowledgements}
This work was performed within the following projects: RÅC (Röntgen-Ångström cluster) “INVISION” project, 2019-2023; EuXFEL R\&D “MHz microscopy at EuXFEL: From demonstration to method”, 2020 - 2022; Horizon Europe EIC Pathfinder “MHz-Tomoscopy” project, 2022-2025; ERC starting grant "3DX-Flash" grant agreement No. 948426. The Slovak co-authors‘ thank grant no.: VEGA - 2/0041/21.

%% Note added by Overleaf: If using bibtex, remove the "references" environment above, and uncomment the following line.
\referencelist{iucr}

     %-------------------------------------------------------------------------
     % TABLES AND FIGURES SHOULD BE INSERTED AFTER THE MAIN BODY OF THE TEXT
     %-------------------------------------------------------------------------

\appendix

\section{Dynamical Theory of X-ray Diffraction}
\label{appendix:Dynamical_Theory_of_X-ray_Diffraction}
We calculate the integrated diffracted intensity $ I_i^{d} $ for symmetric Laue or Bragg diffraction by following the Dynamical Theory of X-ray Diffraction \cite{Authier:book}.

\textbf{Laue Diffraction}  \
The integrated diffraction intensity in Laue geometry $I_{i Laue}^{d}$ is calculated from \cite{Authier:book}~(page 98, equation 4.40) as:
\begin{equation}
    \label{eq:IntegratedIntensityLaue}
    I_{i Laue}^{d}=\frac{\pi \| C_p \chi_h \| \sqrt{\gamma}}{2 sin(2\theta_B)} \int_0^{2\pi \!t/\Lambda_L} J_0(z)dz
    \cdot e^{-\frac{\mu t}{2}(\frac{1}{\gamma_o}+\frac{1}{\gamma_h})}
\end{equation}
where \(C_p\) is the polarization factor, $\chi_h$ is the dielectric susceptibility of the diffraction plane, $\gamma = \gamma_h / \gamma_o$ the asymmetry factor, \(\theta_B\) the Bragg angle, $t$ the thickness of the splitter, $\Lambda_L$ the extinction length in Laue geometry for the specific diffraction plane and asymmetry factor, $J_0(z)$ the zeroth-order Bessel function, $e^{-\mu t(\frac{1}{\gamma_o}+\frac{1}{\gamma_h})}$ the transmission of the diffracted beam with $\mu$ the linear absorption coefficient of the material, $\gamma_o$ and $\gamma_h$ are the direction cosines of incident and diffracted beam relative to the inner normal to the crystal surface.
\begin{equation}
    \label{eq:cosine_o}
\gamma_o = cos(\beta + \theta_B)
\end{equation}
\begin{equation}
    \label{eq:cosine_h}
\gamma_h = cos(\beta - \theta_B)
\end{equation}
where $\beta$ is the asymmetry angle in Laue geometry. 
For symmetric Laue diffraction $\beta=0$, therefore $\gamma_o = \gamma_h$, $\gamma = 1$ and the absorption factor is reduced to $e^{-\mu t cos(\theta_B)}$.

The $ I_i^{d} $ function versus the thickness of the Laue splitters follows Pendellösung oscillations, i.e. the oscillation in intensity between transmitted and diffracted beam due to the symmetric nature of Laue diffraction. 
The maximum $I_{i Laue}^{d}$ is reached when the thickness is comparable to the extinction length, i.e. the length over which virtually all the beam is diffracted.   

\textbf{Bragg Diffraction} \ 
In the Bragg case, there is no exact formula to calculate the $ I_i^{d} $ over a wide range of thicknesses. However, since Bragg diffraction does not present strong Pendellösung oscillations, the integrated diffracted intensity converges rapidly to an average, from \cite{Authier:book}~(page 101, equation 4.43):
\begin{equation}
    \label{eq:IntegratedIntensityBragg}
    I_{i Bragg}^{d}= \frac{8}{3} \| \delta_{os} \| 
\end{equation}
where $\delta_{os}$ is Darwin's width, i.e. half the angular acceptance of the lattice plane for Bragg diffraction: 
% 1x MATH FORMULA Darwin Width
\begin{equation} 
\label{eq:Darwin_width}
 \delta_{os} \, = \, \frac{C_p r_e \lambda^2}{\pi V sin(2 \theta_B)} \sqrt{|\gamma|} \sqrt{F_{c_{hkl}} F_{c_{\bar{h} \bar{k} \bar{l}}}}
\end{equation}
where \(C_p\) is the polarization factor, \(r_e\) is the electron radius, \(\lambda\) is the wavelength, \(V\) the volume of the unit cell, $\gamma = \gamma_h / \gamma_o$, \(F_c\) the structure factor for the particular diffraction plane with Miller's indices \(hkl\) or $\bar{h} \bar{k} \bar{l}$. 
In case of symmetric Bragg geometry, $\beta = \pi / 2$, $\gamma_h = - \gamma_o$, and $|\gamma| = 1$.
The integrated diffracted intensity in Bragg geometry is higher than in Laue geometry because of the reduced  thickness that the diffracted beam must traverse.

\section{Magnification}
\label{appendix:Magnification}
For the recombiners, the asymmetry angle is defined as the angle $\alpha$ between the lattice planes and the physical surface of a crystal. A grazing incidence angle can be used for enlarging the acceptance of the crystal (Eq. \ref{eq:Darwin_width}) \cite{Authier:book} while enlarging the physical size of the diffracted beamlet over the diffraction direction by a magnification factor $M$: 

%1x MATH FORMULA Magnification due to Asymmetry
\begin{equation} 
\label{eq:Magnification}
 M \, = \, \frac{sin\theta_{in}}{sin\theta_{out}} \, = \, \frac{sin(\theta_B + \Delta\theta_{hc} + \alpha)}{sin(\theta_B + \Delta\theta_{oc} - \alpha)}
\end{equation}
where \(\theta_{in}\) and \(\theta_{out}\) are the incident and outgoing angles between the beamlet and the recombiner surface, \(\theta_B\) is the Bragg angle, \(\Delta\theta_{oc}\) and \(\Delta\theta_{hc}\) are correction terms for the incoming and outgoing beam obtained by the dynamical theory of diffraction \cite{Authier:book}, and $\alpha$ is the asymmetry angle. 
In our setup, enlarging the beamlet's physical size can be beneficial since the beamlet was already shrunk due to the asymmetry of the splitter, by a factor of 0.818 for both the C113 and Si113 splitters. 
The total magnification of the beamlet is attained by multiplying the magnifications produced by the splitter and the recombiner. 
Therefore, we can select a recombiner asymmetry that increases the acceptance while making the shape of the beamlet more symmetric, or similar to the shape of the field of view of the camera. 
Cameras often have a larger horizontal field of view. 
As an example, the MHz camera Shimadzu HPV-X2 used in this study has a field of view of 400 horizontal (H) $\times$ 250 vertical (V) pixels, so an aspect ratio H/V of 1.6. 
Therefore, we can adjust the magnification to approximate this value to have beamlets that fit the field of view of the camera. 
In our specific case, this optimization leads to selecting a grazing asymmetry of 10° because the magnification factor for the selected {110} planes (Table \ref{tab:In-Parallel_recombiner_families}) re-balances the shrinking caused by the splitter and produces beamlets with an aspect ratio similar to the field of view of the camera. 
For a germanium recombiner with 10° asymmetry, at 19.1 keV the magnification is 3.11, 1.93, 1.52 for (440), (660), and (880) diffraction planes respectively so resulting in a total magnification of the image of 2.55, 1.58, 1.24. 
At 12.55 keV the magnification is 1.90 and 1.38 for (440) and (660) diffraction planes, respectively, resulting in a total magnification of the image of 1.56 and 1.13.

\section{Mechatronics}
\label{appendix:Mechatronics}

\subsection{Motor assembly} 
\label{appendix:Mechatronics:motor_assembly}
The low acceptance of some of the crystal optics, in particular the recombiners of the In-Parallel geometry, calls for very precise and stable mechanics for beam alignment and keeping the alignment stable over the duration of an experiment. For this purpose, we developed together with SmarAct GmbH precise 6-axis positioners composed of 6 stacked motors. The order of arrangement is important since the positioners must be able to align the crystal lattice planes with the rotation angle controlling the Bragg angle. This is particularly important for the In-Parallel splitter since it must meet multiple diffraction conditions, so two orthogonal rotation axes must be functionally independent. Consider a reference system with a horizontal x-axis, a vertical y-axis, and a z-axis aligned in beam direction. Alpha, beta, and gamma are the rotation angles around these axes respectively. All the positioners require the same base platform composed of 5 motors, from bottom to top: 2 linear horizontal axes (XZ), a vertical linear axis (Y), a rotation around the vertical axis (beta), and a tilt. The final motor of the positioner varies depending on the specific optics it will be used with, such as a recombiner, a splitter In-Parallel mode, beta In-Line mode.

\begin{figure}
    \includegraphics[width=1.0\textwidth]{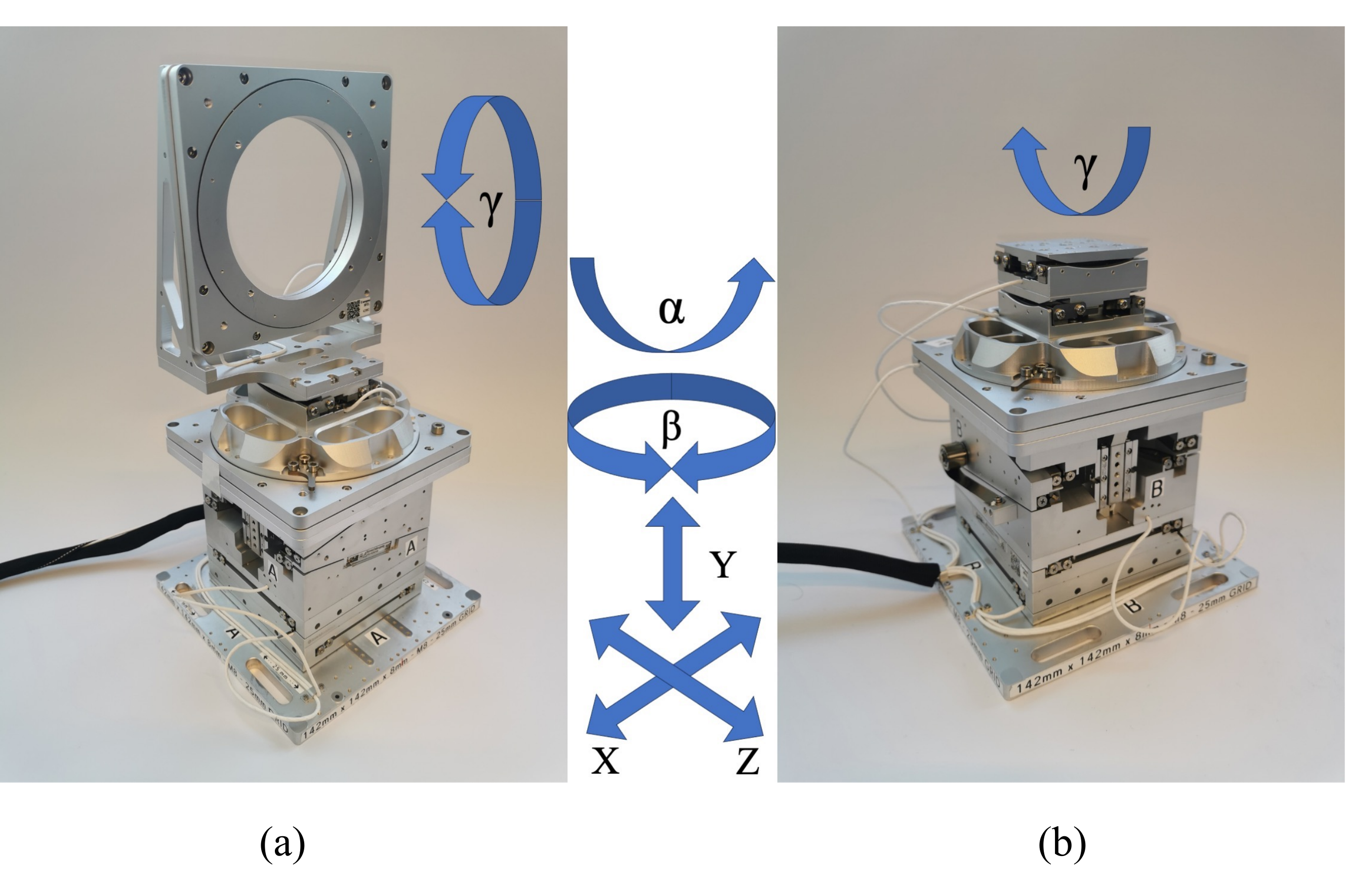}
    \label{fig:towers}
    \caption{Picture of two models of the 6-axes piezo positioners used for aligning the crystals. For both models shown in (a) and (b), the first five motors are identical. From the bottom up these are: two horizontal linear motors (X, Z), a vertical motor (Y), a rotation around the vertical axis ($\beta$), and a tilt ($\alpha$). The top motor ($\gamma$) is a rotary motor in (a) and a tilt in (b). (a) is used as positioner A and C, while (b) is used as positioner B (sections \ref{subsubsection:In-Parallel_Geometry}, \ref{subsubsection:In-Line_Geometry}).}  
\end{figure}

\subsubsection{In-Parallel Geometry:} \label{subsubsection:In-Parallel_Geometry} The positioners can be divided into 3 types according to the function of the mounted optics.

A - Laue Splitter Positioner (Fig. \ref{fig:towers}a): The beta rotation aligns the Bragg angle of the beamlets in the horizontal plane, the alpha tilt aligns the Bragg angle of the vertical beamlets, and the gamma rotation aligns their angle around the beam. The adjustments in alpha and beta are critical since they control the Bragg angle, therefore any misalignment in these angles could cause the splitter to go out of diffraction. 

B - Horizontal Recombiner Positioner (Fig. \ref{fig:towers}b): The Bragg angle is regulated by the beta rotation stage. In addition to the rotation around angle beta, two additional tilts around alpha and gamma are required to adjust the diffracted beamlets. 

C - Vertical Recombiner Positioner (Fig. \ref{fig:towers}a): These positioners are identical to type A but rotated by 90° around the vertical axis. In order to align the beamlet in the vertical direction, the positioners of type C affect a tilt along gamma. In this case, the Bragg angle is regulated by the top alpha rotation stage, also used for aligning the recombiner to different diffraction orders, e.g. (220), (440), (660).

\subsubsection{In-Line Geometry:} \label{subsubsection:In-Line_Geometry} In this geometry, all optical components diffract a single beamlet over the same diffraction plane. However, two different positioner types are needed in case different splitters are used. In both types, the Bragg angle is regulated by the beta rotation stage. 

A - Skew Planes Positioner (Fig. \ref{fig:towers}a): This positioner is identical to type "A - Splitter Positioner". It is used to align skewed planes, i.e. planes non-parallel to any of the sides of the splitter. The large gamma rotation aligns the skewed plane with the horizontal diffraction plane. As for the previous positioner type, if the diffraction plane is vertical, it is necessary to flip the entire assembly composed of the top three rotary motors by 90°.

B - Standard In-Line Positioner (Fig. \ref{fig:towers}b): This type of positioner is identical to type "B - Horizontal Recombiner Positioner". The two top tilts around alpha and gamma fine-tune the alignment of the diffraction plane in the horizontal plane. In our experiments, we used a horizontal diffraction plane. However, these positioners can also be utilized for a vertical diffraction plane by rotating the top three rotary motors by 90°. This can be achieved either by using a right-angle bracket or by employing a type C positioner.

\subsubsection{Clamping:} 
\begin{figure}
\includegraphics[width=1\textwidth]{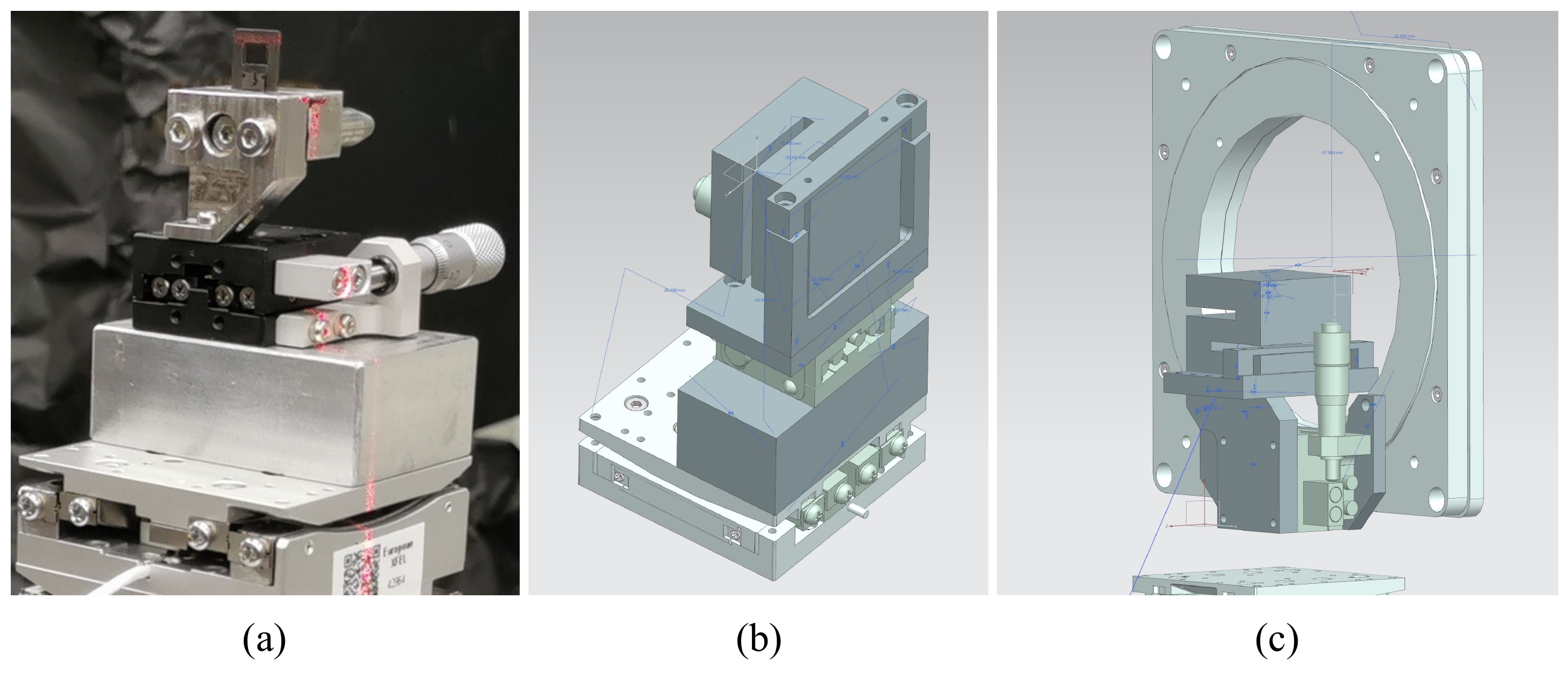}
\label{fig:manual_stage}
\caption{(a) Picture of a beam splitter being pre-aligned to the beam with a laser and the manual stage on top of the 6-axis piezo positioners. (b) CAD drawing of the horizontal recombiner holder. (c) CAD drawing of the vertical recombiner holder.}  
\end{figure}

In addition to the piezo motor structure described above, custom holders were designed to clamp the crystals and align their diffracting planes in the center of rotation of the positioners with micrometric precision (Fig. \ref{fig:manual_stage}). For the splitters, the crystal's center is placed in the center of the rotation, while for the recombiners, it is the center of the main face. To ensure that the crystals (Fig. \ref{fig:crystals}) are clamped without experiencing stress in their optic area, stress relief cuts were incorporated \cite{Liubov:2019}. In this design, the clamping occurs on the side opposite the optic surface, while deep cuts separate the clamped portion from the optic surface. The stress resulting from clamping is distributed in the material within the stress relief cuts, which is the thinnest and longest part of the crystal. Consequently, any deformation occurs in this specific region. This deformation causes a net rotation of the optic part, but it does not introduce any curvature or other deformations to the optic part itself. The rotary motors of the 6-axis positioner can easily compensate for this net rotation. 

% SmarAct Stability Tests 
\subsection{Stability and repeatability tests} 
\label{appendix:Mechatronics:Stability}

\begin{figure}
\label{fig:interferometric_measure}
\includegraphics[width=1.0\textwidth]{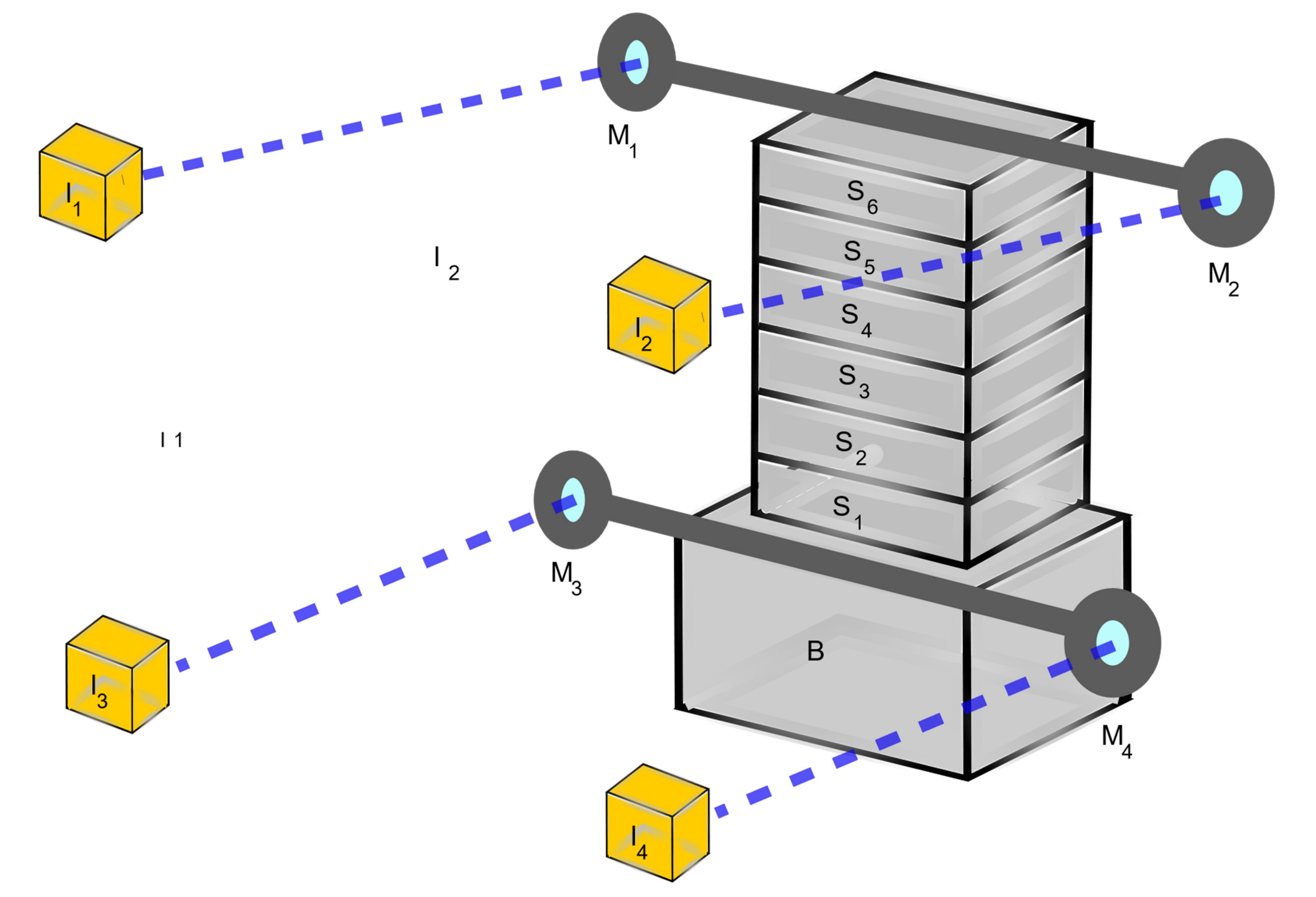}
\caption{Sketch of the interferometric setup used for the stability and repeatability measures on the 6-axis piezo positioners, with $I_n$ the interferometric units, $M_n$ the mirrors, $S_n$ the motors, and $B$ the base.
Two bars with mirrors at the end were affixed to the top and the bottom of the positioner. 
The difference in position between the two mirrors at the ends of one bar gives the rotary angle, which controls the Bragg angle. 
}
\end{figure}

\begin{figure}
    \label{fig:stability_complete}   
    \includegraphics[width=0.5\textwidth]{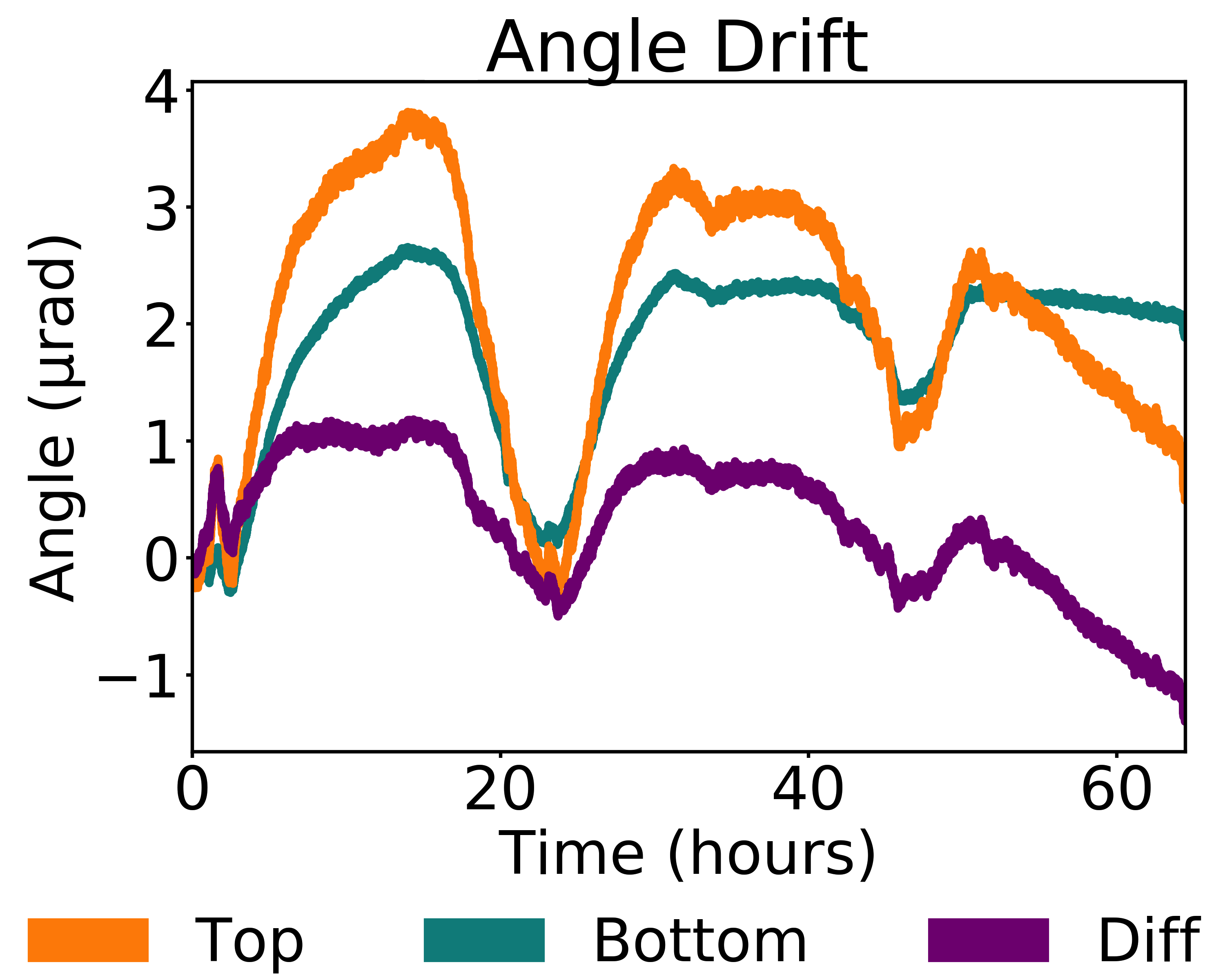}
\caption{Stability tests on the 6-axis piezo positioners over 64 hours. 
Drift of the rotary angle (Bragg angle) at the top, bottom and difference of the two, the latter representing the real drift of the rotary angle when the stability of the structure under the 6-axis positioner is eliminated. 
        }  
\end{figure}

\begin{figure}
    \label{fig:environmental_data}   
    \includegraphics[width=0.5\textwidth]{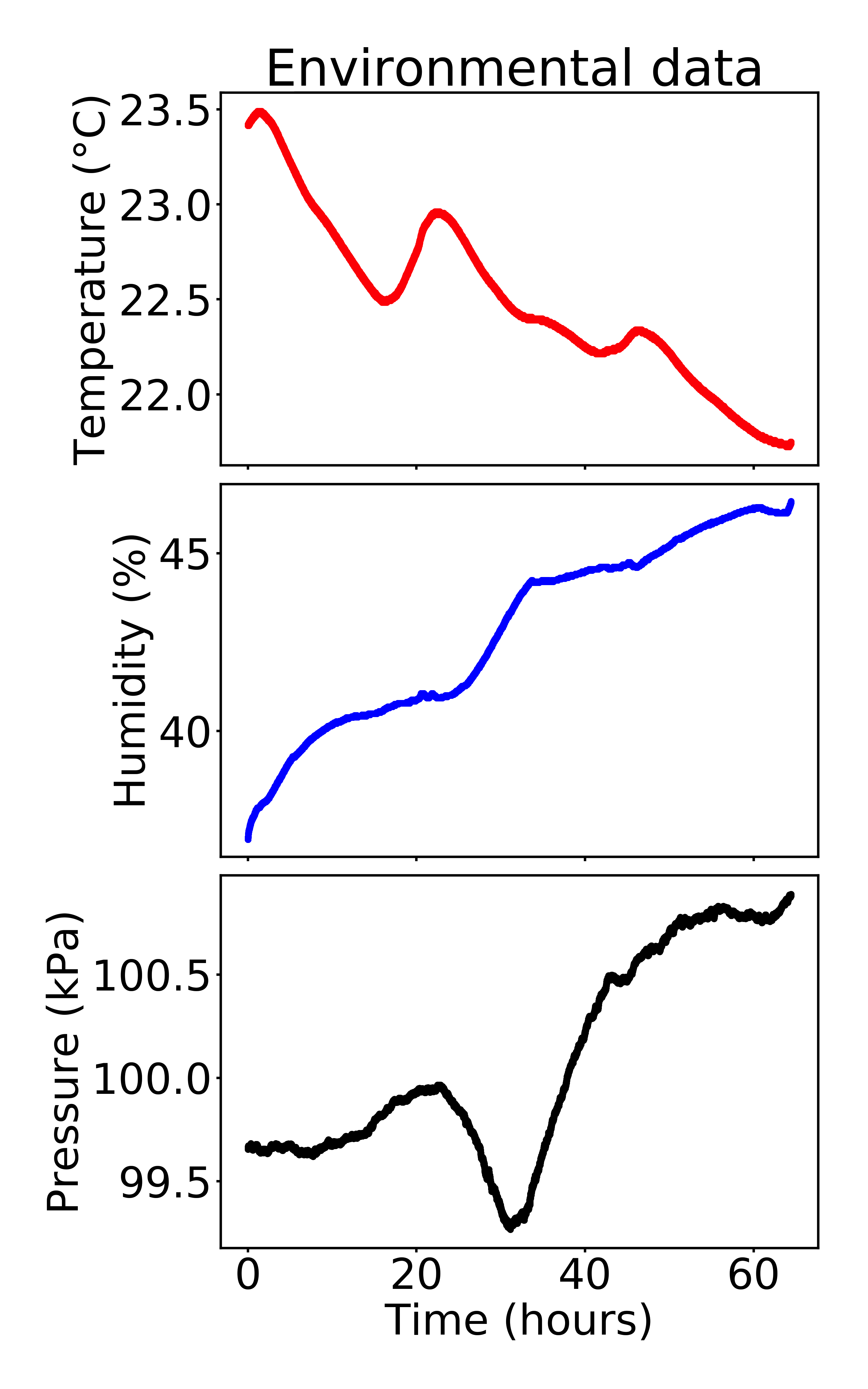}
\caption{Environmental data during the stability tests on the 6-axis piezo positioners, spanning over 64 hours. 
        }  
\end{figure}

Analysis of stability and repeatability were performed by interferometric measurement with 20 nrad resolution. The interferometric system was composed of two "Picoscale" interferometers coupled with an aluminum bar with mirrors at the extremities, so the angular displacement is calculated knowing the length of the bar and the movement of the mirrors. One interferometric system was mounted on top of a piezo 6-axis positioner (Fig. \ref{fig:interferometric_measure}), while another was mounted on its base. Temperature, humidity, and air pressure were monitored in the room during the measurement. The stability of the system was measured over intervals spanning a maximum of 64 hours. During this measure, environmental data as temperature, pressure and humidity were measured (Fig. \ref{fig:environmental_data}). The stability of the 6-axis positioner resulted in less than 3 µrad over 64 hours (Fig. \ref{fig:stability_complete}), which gives sufficiently stable conditions to align crystalline optics from the simulations (Section \ref{subsubsection:DuMond_diagrams}). The repeatability of the system was also tested (Table \ref{tab:repeatability}) by moving over a range of positions, with step size and range size changing. The maximum reversal error for the largest range of 174 µrad was within 230 nrad (1 sigma variation), so highly reproducible. 

\begin{table}
\label{tab:repeatability}
\begin{tabular}{lccc}      % Alignment for each cell: l=left, c=center, r=right
                     &            &\underline{Travel range:}  \\
 Criterion           & 1.74 µrad  & 17.4 µrad  & 174 µrad  \\
\hline
Max. Reversal Error (nrad)  & 30.39   & 107.32 & 230.97  \\
Unidirectional Repeatability (nrad) & 49.47  & 50.11  & 178.32  \\
Bidirectional Repeatability (nrad)  & 50.06  & 70.38  & 200.5   \\
 \\

\end{tabular}

\caption{Results of the repeatability tests of the 6-axis positioners, performed by moving the rotary motor in steps over a range of positions, with step size and range size changing. All the repeatability data are for 1 sigma, 10 target positions, and 25 repetitions for each target position. 
}
\end{table}

% ESRF Topography 
\section{X-ray diffraction imaging}
\label{appendix:Topography}
\begin{figure}
\label{fig:diamond_topography}
\includegraphics[width=1.0\textwidth]{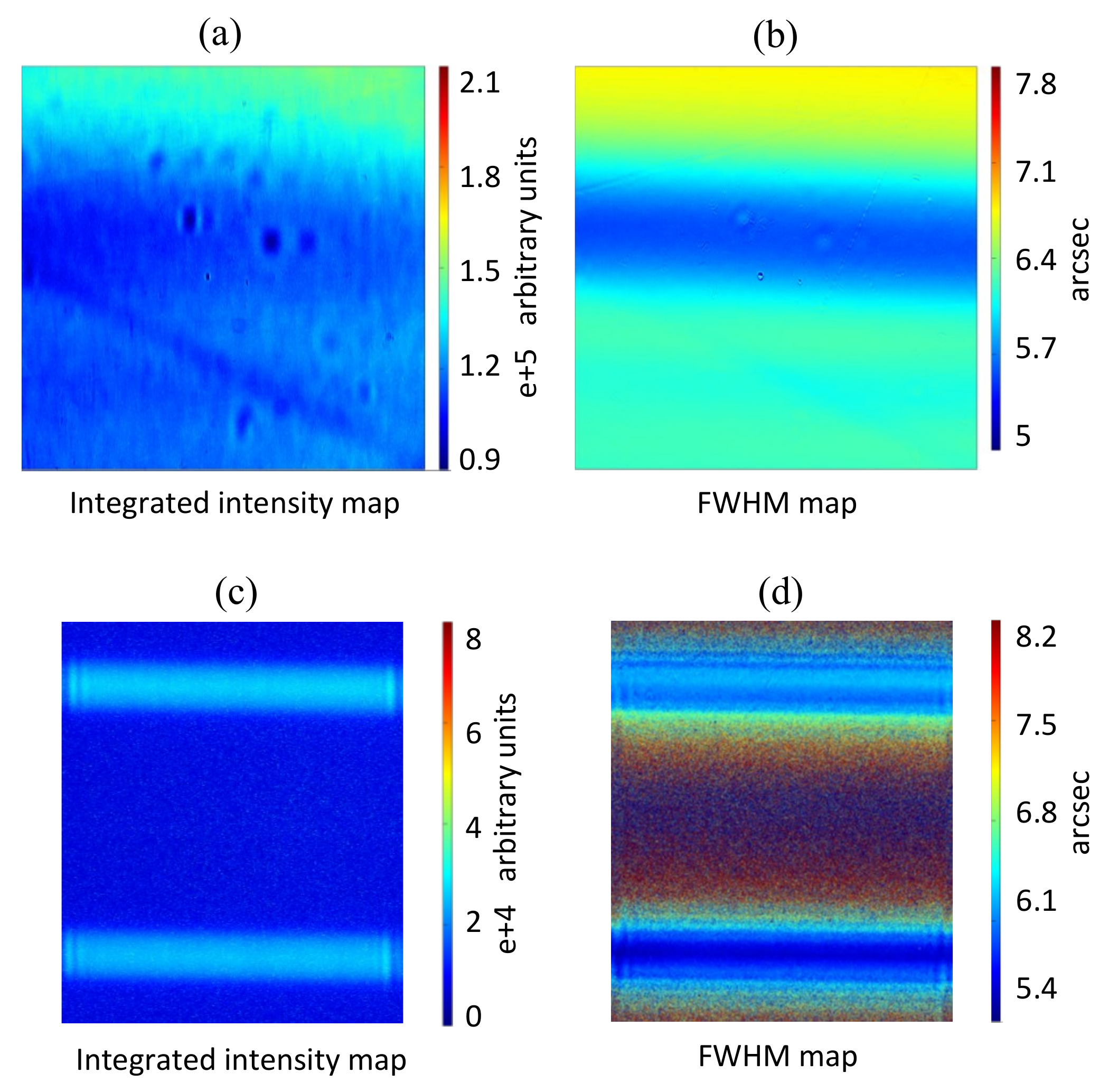}
\caption{Monochromatic high-resolution X-ray topography at BM05 ESRF beamline on a diamond beam-splitter. The analysis was conducted via Laue diffraction on its (220) lattice planes, both on the (220) main face and the (2-20) orthogonal to the main face. The photon energy is 20 keV, the field of view 1.3 mm $\times$ 1.3 mm, and the pixel size 0.65 µm. 
(a,b) are images of the crystal surface, image size 1.3 mm $\times$ 1.3 mm. 
(a) Integrated intensity map of the surface, i.e. map of the intensity diffracted by each point on the surface for a specific Bragg angle of the splitter. (b) FWHM map of the surface, i.e. map of the diffraction passband for each point on the surface. 
(c,d) are zoomed images of two crystal sections, the distance between the sections being 500~µm.
(c) Section topography, integrated intensity map for sections of the splitter. (d) Section topography, FWHM map for sections of the splitter.
}
\end{figure}

The quality of the crystals was analyzed by high-resolution monochromatic X-ray topography at the ESRF beamline BM05 \cite{TranThi:vc5001, BM05:2004}. This includes rocking curve imaging and section topography to investigate both the surface and the bulk of the crystals. The beamline was set to a 20 keV monochromatic beam after a double-diffraction (111) silicon monochromator. Different diffraction planes were analyzed to test the crystal quality for dislocations that can appear over particular directions (Fig. \ref{fig:diamond_topography}). These images were taken with a high-resolution detector with a field of view of 1.3 mm $\times$ 1.3 mm and a pixel size of 0.65 µm, stitching the entire optic surface of the crystals. The diamond splitters (shown in Fig. \ref{fig:diamond_topography}) performed well during rocking curve imaging, with good crystalline quality and FWHM 5.5 arcsecs. A map of the optic surface also shows some polishing waves, within tolerance. Section scans were performed by taking three sections from one surface to the opposite surface so to analyze the bulk of the crystal, the sections were spaced by a distance of 400 µm each and 1.3 mm wide. Section scans show no defects or inclusions on the surface or in the bulk. The topography of the recombiners shows a different picture. Germanium recombiners appear to have a rougher surface, even if the quality is uniform and consistent over the whole sample. In this case, section scans are not possible because of the thickness of the sample. This rougher surface can be attributed to the brittle structure of germanium and the less-developed finishing technologies compared to silicon or diamond. 

\begin{figure}
\label{fig:germanium_topography}
\includegraphics[width=1.0\textwidth]{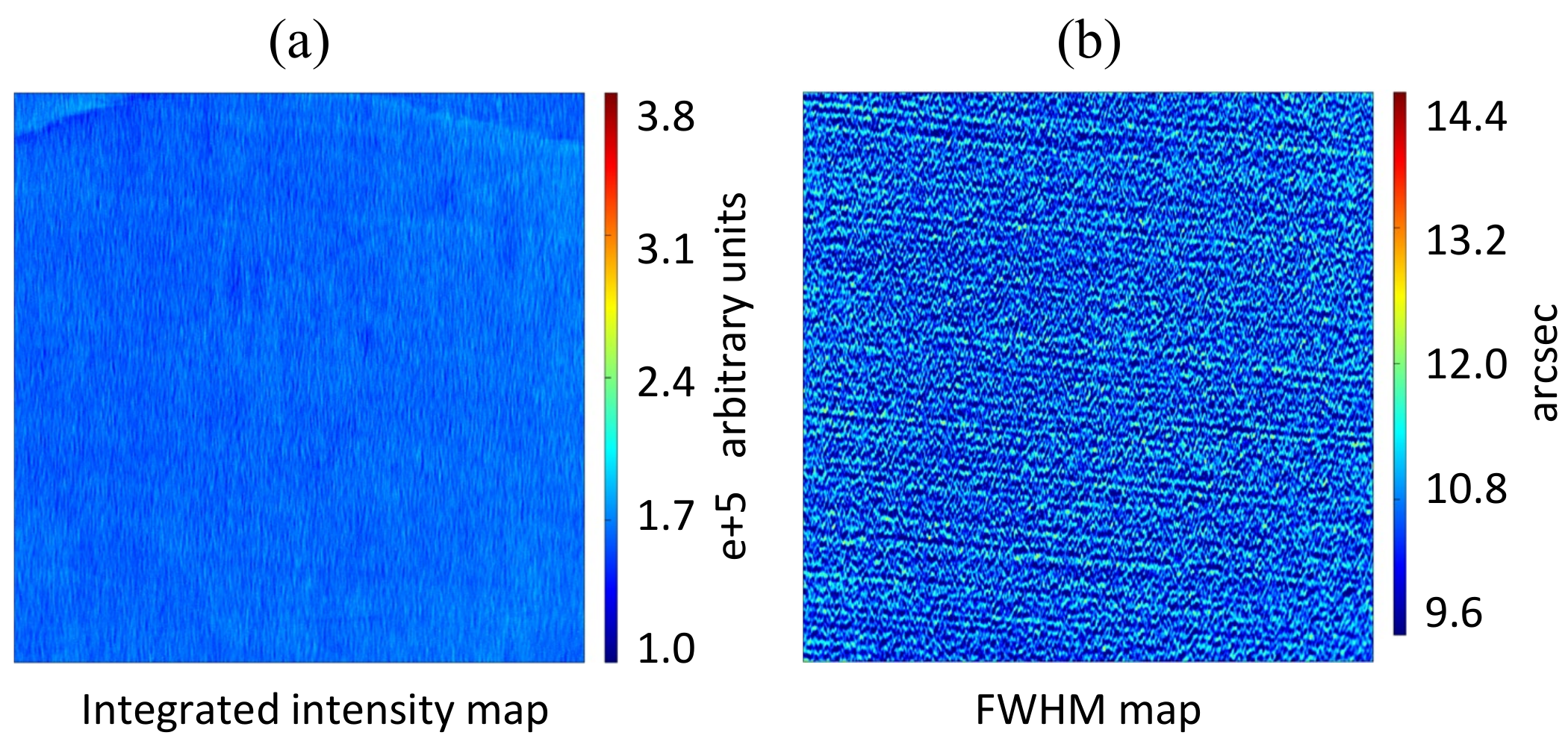}
\caption{Monochromatic high-resolution X-ray topography at BM05 ESRF beamline for the In-Parallel geometry germanium recombiners. The analysis was conducted via Bragg diffraction on the recombiners' (440) lattice planes. The images represent an area of the crystal surface 1.3 mm wide horizontally and 5 mm long vertically due to the elongated footprint in the direction of diffraction. The photon energy is 20 keV, the field of view 1.3 mm $\times$ 1.3 mm, and the pixel size 0.65 µm. (a) Integrated intensity map of the surface, i.e. map of the intensity diffracted by each point on the surface for a specific Bragg angle of the recombiner. (b) FWHM map of the surface, i.e. map of the diffraction passband for each point on the surface.
}
\end{figure}

\end{document}